\documentclass{elsart}
\usepackage{graphicx}

\def\deg{$^{\circ}$}
\def\micron{$\mu$m}
\def\mathrelfun#1#2{\lower3.6pt\vbox{\baselineskip0pt\lineskip.9pt
  \ialign{$\mathsurround=0pt#1\hfil##\hfil$\crcr#2\crcr\sim\crcr}}}
 
\def\simgt{\mathrel{\mathpalette\mathrelfun >}}

\def\fun#1#2{\lower3.6pt\vbox{\baselineskip0pt\lineskip.9pt
  \ialign{$\mathsurround=0pt#1\hfil##\hfil$\crcr#2\crcr\sim\crcr}}}
\def\zu{\rm\,}     % units in formulae are roman typed
\def\etal{{\it et~al. }}

\begin{document}

\begin{frontmatter}
\runauthor{A. Beno{\^\i}t \etal}
%   \thesaurus{02         % A&A Section 2: Cosmology
%   \thesaurus{02         % A&A Section 2: Cosmology
%     (03.02.1;  % Balloons
%      03.09.4;  % Instrumentation: photometers
%      13.20.1;  % Submillimeter
%      12.03.1;  % Cosmology: cosmic microwave background
%      12.03.3)  % Cosmology: observations,
%            }
\title{Archeops: A High Resolution,  Large Sky Coverage 
Balloon Experiment for Mapping CMB Anisotropies}
%\authorrunning{A. Beno{\^\i}t \etal}
%\titlerunning{Archeops for Mapping CMB Anisotropies}

% \subtitle{}

\author[Inst1]{ A.~Beno{\^\i}t}, 
\author[Inst2]{ P.~Ade},
\author[Inst3]{ A.~Amblard},
\author[Inst4]{ R.~Ansari},
\author[Inst5]{ E.~Aubourg},
\author[Inst6]{ J.~Bartlett},
\author[Inst7]{ J.--P.~Bernard},
\author[Inst8]{ R.~S.~Bhatia},
\author[Inst6]{ A.~Blanchard},
\author[Inst9]{ J.~J.~Bock},
\author[Inst10]{ A.~Boscaleri},
\author[Inst11]{ F.~R.~Bouchet},
\author[Inst4]{ A.~Bourrachot},
\author[Inst12]{ P.~Camus},
\author[Inst4]{ F.~Couchot},
\author[Inst13]{ P.~de~Bernardis},
\author[Inst3]{ J.~Delabrouille},
\author[Inst14]{ F.--X.~D{\'e}sert},
\author[Inst11]{ O.~Dor{\'e}},
\author[Inst6]{ M.~Douspis},
\author[Inst12]{ L.~Dumoulin},
\author[Inst15]{ X.~Dupac},
\author[Inst16]{ P.~Filliatre},
\author[Inst3]{ K.~Ganga},
\author[Inst2]{ F.~Gannaway},
\author[Inst1]{ B.~Gautier},
\author[Inst15]{ M.~Giard},
\author[Inst3]{ Y.~Giraud--H{\'e}raud},
\author[Inst7]{ R.~Gispert$^\dag$\thanksref{RG} },
\author[Inst3]{ L.~Guglielmi},
\author[Inst3]{ J.--C.~Hamilton},
\author[Inst17]{ S.~Hanany},
\author[Inst4]{ S.~Henrot--Versill{\'e}},
\author[Inst8]{V.~V.~Hristov},
\author[Inst3]{ J.~Kaplan},
\author[Inst7]{ G.~Lagache},
\author[Inst7]{ J.--M.~Lamarre},
\author[Inst8]{ A.~E.~Lange},
\author[Inst1]{ K.~Madet},
\author[Inst2]{ B.~Maffei},
\author[Inst17]{ D.~Marrone},
\author[Inst13]{ S.~Masi},
\author[Inst18]{ J.~A.~Murphy},
\author[Inst16]{ F.~Naraghi},
\author[Inst13]{ F.~Nati},
\author[Inst16]{ G.~Perrin},
\author[Inst7]{ M.~Piat},
\author[Inst7]{ J.--L.~Puget},
\author[Inst16]{ D.~Santos},
\author[Inst2]{ R.~V.~Sudiwala},
\author[Inst3]{ J.--C.~Vanel},
\author[Inst11]{ D.~Vibert},
\author[Inst2]{ E.~Wakui},
\author[Inst5]{ D.~Yvon}

\thanks[RG]{Richard Gispert passed away few weeks after his return
from the mission to Trapani}

\address[Inst1]{
Centre de Recherche sur les Tr\`es Basses Tem\-p\'e\-ra\-tures, 25 Avenue
des Martyrs BP166, F--38042 Grenoble Cedex 9, France}
\address[Inst2]{
Queen Mary and Westfield College, London, UK}
\address[Inst3]{
Physique Corpusculaire et Cosmologie, College de
France,  11 pl. Marcelin Berthelot, F-75231 Paris Cedex 5, France}
\address[Inst4]{
Laboratoire de l'Acc{\'e}l{\'e}rateur Lin{\'e}aire, BP~34, Campus
Orsay, 91898 Orsay Cedex, France}
\address[Inst5]{
SPP/DAPNIA/DSM, CEA-Saclay, F--91191 Gif-sur-Yvette  Cedex, France}
\address[Inst6]{
Observatoire de Midi-Pyr{\'e}n{\'e}es, 
14 Avenue E. Belin, 31400 Toulouse, France}
\address[Inst7]{
Institut d'Astrophysique Spatiale, B\^at.  121, Universit\'e Paris XI,
F--91405 Orsay Cedex, France}
\address[Inst8]{
California Institute of Technology, Pasadena, CA, USA}
\address[Inst9]{
Jet Propulsion Laboratory, Pasadena, CA, USA}
\address[Inst10]{
IROE--CNR, Firenze, Italy}
\address[Inst11]{
Institut d'Astrophysique de Paris, 98bis, Boulevard Arago, 75014 Paris, France}
\address[Inst12]{
CSNSM--IN2P3, B{\^a}t 108, 91405 Orsay Campus, France}
\address[Inst13]{
Gruppo di Cosmologia Sperimentale, Dipartimento di Fisica, Universita
``La Sapienza'', P. A. Moro, 2, 00185 Roma, Italy}
\address[Inst14]{
Laboratoire d'Astrophysique, Observatoire de Grenoble BP 53, 414 rue
de la piscine, F--38041 Grenoble Cedex 9, France }
\address[Inst15]{
Centre d'\'Etude Spatiale des Rayonnements, 9 avenue du Colonel Roche, 
BP 4346, F--31028 Toulouse Cedex 4, France}
\address[Inst16]{
Institut des Sciences Nucl{\'e}aires, 53 Avenue des Martyrs, 38026
Grenoble Cedex, France}
\address[Inst17]{
School of Physics and Astronomy, 116 Church St. S.E., University of
Minnesota, Minneapolis MN 55455, USA}
\address[Inst18]{
Experimental Physics, National University of Ireland, Maynooth, Ireland}

%    \offprints{A. Beno{\^\i}t, benoit@polycnrs-gre.fr} 
\date{~May 22, 2001: Final accepted version for AstroParticle Physics}

\begin{abstract}
     Archeops is a balloon--borne instrument dedicated to measuring
cosmic microwave background (CMB) temperature anisotropies at high
angular resolution ($\sim 8$ arcminutes) over a large fraction ($\sim
25$\%) of the sky in the millimetre domain.  Based on Planck High
Frequency Instrument (HFI) technology, cooled bolometers (0.1~K) scan
the sky in total power mode with large circles at constant elevation.
During the course of a 24--hour Arctic--night balloon flight, Archeops
will observe a complete annulus on the sky in four frequency bands
centered at 143, 217, 353 and 545~GHz with an expected sensitivity to
CMB fluctuations of $\sim 100\; \mu$K for each of the 90 thousand 20
arcminute average pixels.  We describe the instrument and its
performance obtained during a test flight from Trapani (Sicily) to
Spain in July 1999.

%\keywords{Balloon -- Instrumentation: photometers -- Submillimeter --
%   Cosmology: cosmic microwave background -- Cosmology: observations}
%   \thesaurus{02         % A&A Section 2: Cosmology
%     (03.02.1;  % Balloons
%      03.09.4;  % Instrumentation: photometers
%      13.20.1;  % Submillimeter
%      12.03.1;  % Cosmology: cosmic microwave background
%      12.03.3)  % Cosmology: observations,
%            }
\end{abstract}
\end{frontmatter} 
%________________________________________________________________
\clearpage

\section{Introduction}

The anisotropies of the cosmic microwave background (CMB) are a
goldmine of cosmological information. They tell us about the state of
the Universe at the tender age of a few hundred thousand years
(redshift $\sim 1000$) when the primeval plasma recombined and
unveiled the small fluctuations that would eventually form the
large--scale structure of the present--day Universe.  The
applicability of linear perturbation theory to these small
inhomogeneities provides the cosmologist with a relatively direct link
between observable and theoretical quantities over a large range of
physical scales.  Accurate measurements of the anisotropies can
therefore be used to constrain fundamental parameters of the standard
Big Bang cosmogoly, at least within the adopted theoretical framework
(e.g., Inflation).  For the case of inflation--generated
perturbations, this approach most directly constrains the spatial
geometry of the Universe, the physical particle densities (such as the
baryon or dark matter density) and the form of the primeval
perturbation spectrum (scalar and tensor modes), as well as the
ionization history since recombination.  A large international effort
has set as its goal high precision measurements of the CMB
anisotropies, employing numerous ground--based, balloon and
space--borne projects; among the latter MAP to be launched by NASA in
2001, and Planck to be launched in 2007 by ESA.

\begin{figure}
\resizebox{\hsize}{!}{\includegraphics{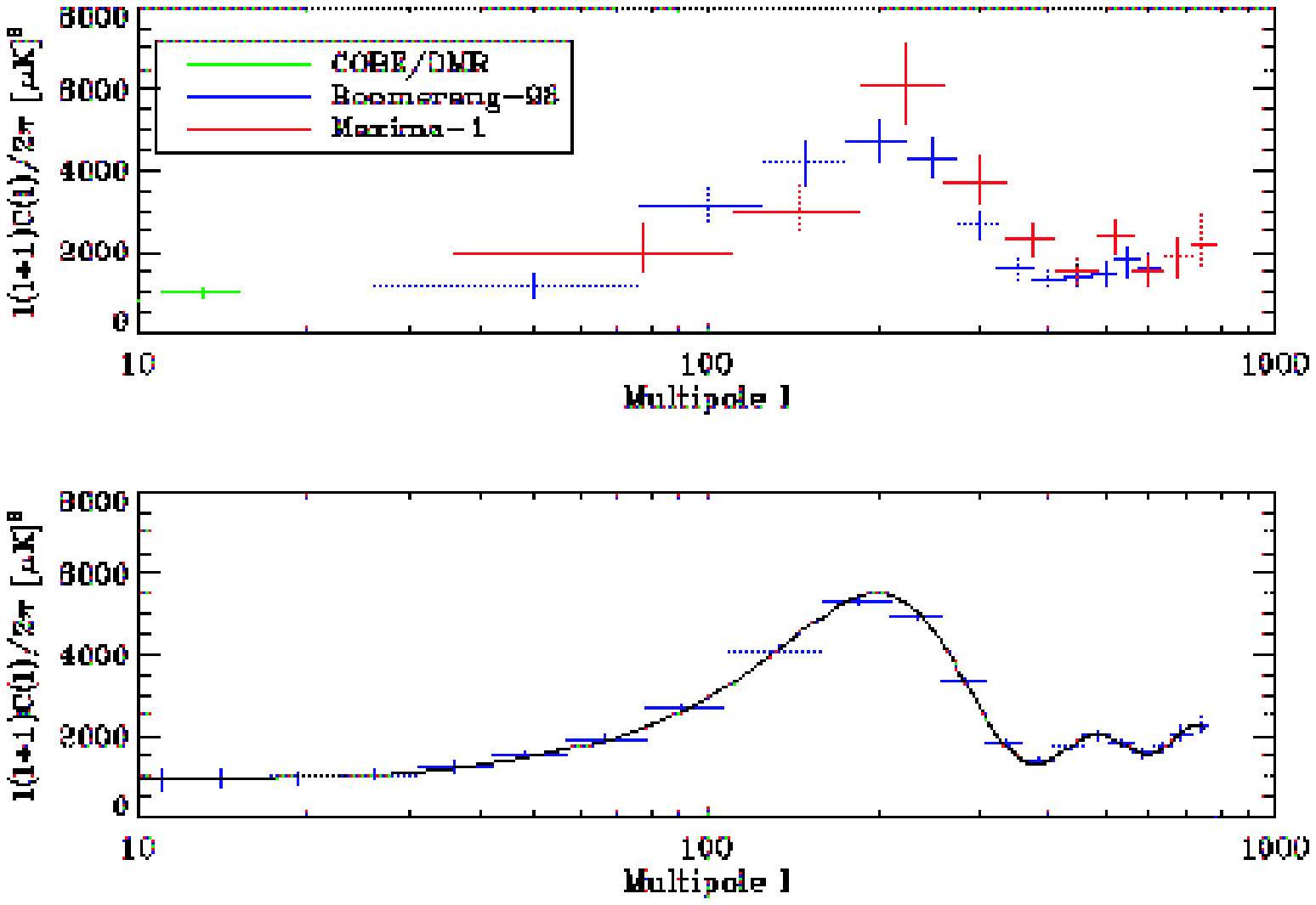}}
\caption{The expected accuracy with which the CMB temperature power spectrum
could be determined using a 24-hour data from Archeops. Top:
COBE/DMR, B98 and Maxima-1 independent band powers
\cite{Bennett:1996}, \cite{Bernardis:2000}, \cite{Hanany:2000}. Bottom: the
COBE/DMR, B98 and Maxima-1 joint analysis best fit with weak prior (solid
line) \cite{Jaffe:2000} and expected band powers from  
Archeops (data points) in bins that have a constant $\Delta ln \ell=0.375$ below $\ell = 107$
and a constant $\Delta \ell = 50$ at higher $\ell$'s. 
A 24-hour Archeops data set should
provide an important overlap between COBE data and current higher
resolution measurements, thus giving a strong determination of the absolute
amplitude of the first acoustic peak. The expected error bars do not
include systematic and calibration uncertainties or 
the effects of removing foregrounds. We assumed a sky coverage of 
25\% and nominal sensitivities (a factor 2 better than 
measured during the Trapani flight) for 8 bolometers at 143 GHz
and 8 at 217 GHz.}\label{fig:sensi}
\end{figure}

     Archeops is a balloon--borne instrument based on Planck High
Frequency Instrument (HFI) technology that is optimized to cover a
large fraction of sky at high angular resolution.  The use of cold
bolometers at the focus of a telescope with an aperture of 1.3~meter
allows the HFI to reach an angular resolution better than 10~arcmin at
millimeter wavelengths.  The goal for Archeops is to map $\sim 25$\%
of the sky with 8~arcmin resolution in four frequency bands -- 143,
217, 353 and 545~GHz -- to obtain a limiting CMB sensitivity of
$\Delta T/T \simeq 3\times 10^{-5}$ per coadded 20~arcmin sky pixel.
This is achieved by spinning the payload ($\sim 2-3$~rpm) so that the
detectors scan the sky at a constant elevation of $\sim 41$\deg,
mapping out a large annulus on the sky during a 24--hour flight.  The
angular resolution is achieved with a 1.5~meter aperture telescope,
and sensitivity is obtained with 8 bolometers at 143~GHz, 8 at
217~GHz, 6 at 353~GHz, and 2 at 545~GHz. The CMB signal is recorded by
the 143 and 217~GHz detectors, and the 353 and 545~GHz detectors
monitor atmospheric and interstellar dust emission.  A 24-hour
night-time balloon flight is possible during the Arctic Winter from
the Swedish Esrange Station (near Kiruna on the Arctic Circle)
operated by the French Centre National d'Etudes Spatiales (CNES).
Comparison in Fig.~\ref{fig:sensi} with the current observations
\cite{Bennett:1996}, \cite{Bernardis:2000}, \cite{Hanany:2000} shows
the potential of this type of mission and in particular its ability
to probe a broad range of angular scales. Namely, it should not only
give new measurements with an angular resolution high enough to probe
the second peak angular domain, but its measurement of large angular
scales fluctuations should also allow an absolute measurement of the
height of the first peak. Indeed, by connecting for the first time in
a single experiment large scale fluctuations as measured by COBE/DMR
to contemporary medium scales measurements, it should offer an
absolute calibration of the first "acoustic" peak height, an issue
still unclear even after the BOOMERanG and MAXIMA results. Note that
both these issues should be confirmed too by the TOPHAT balloon
experiment \cite{Martin:1996} and MAP satellite. Timescale for
Archeops experiment is to have two campaigns in the Winter 2000 and
Winter 2001. Additional information and updates may be found at our
web--site\footnote{ {\tt http://www.archeops.org }}.

In this paper we describe in detail the instrument and its performance
during its test flight from the Italian Space Agency's base in Trapani
(Sicily) in July 1999.  The next section presents the instrument
itself.  This is followed in Section 3 by a discussion of the ground
calibration befo re flight.  General aspects of the test flight are
given in Section 4, followed by a description of the in--flight
performance (Sec.~5) and calibration (Sec.~6).  Section 7 gives a
brief conclusion.

\section{Hardware Description}

     In this section we describe in detail the Archeops optics,
detectors, electronics and flight package, including a star sensor for
pointing reconstruction.

\subsection{Telescope}
The Archeops telescope is a two mirror, off-axis, tilted Gregorian
telescope consisting of a parabolic primary and an elliptical
secondary.  An off-axis Gregorian system has higher aperture
efficiency and lower sidelobe response compared to an equivalent
f-number on-axis system \cite{Dragone:1974}.  The telescope
satisfies the Mizuguchi-Dragone condition
\cite{Mizuguchi:1978}, \cite{Dragone:1982} in which astigmatism is canceled
to first order and there is no cross-polarization in the center of the
field of view.  Cross-polarization is also small away from the center
of the field.

The telescope was designed to provide diffraction--limited performance
when coupled to single mode horns producing beams with FWHM of 8
arcminutes or less at frequencies higher than $\sim 140$~GHz. The target beam
sizes, wavelengths and edge taper are similar to those designed for
the HFI on board the Planck satellite\footnote{Thus, the optical
design is similar to the one proposed for Planck during phase-A of the
project except that the Archeops telescope has lower edge taper, because of the
significantly larger primary mirror, and better optical
performance.}. Within a circular radius of 1.8\deg\ from the center
of the field--of--view, the telescope has a wave-front error of less
than 6\% at all wavelengths.

\begin{figure}[t]
\resizebox{\hsize}{!}{\includegraphics[angle=270]{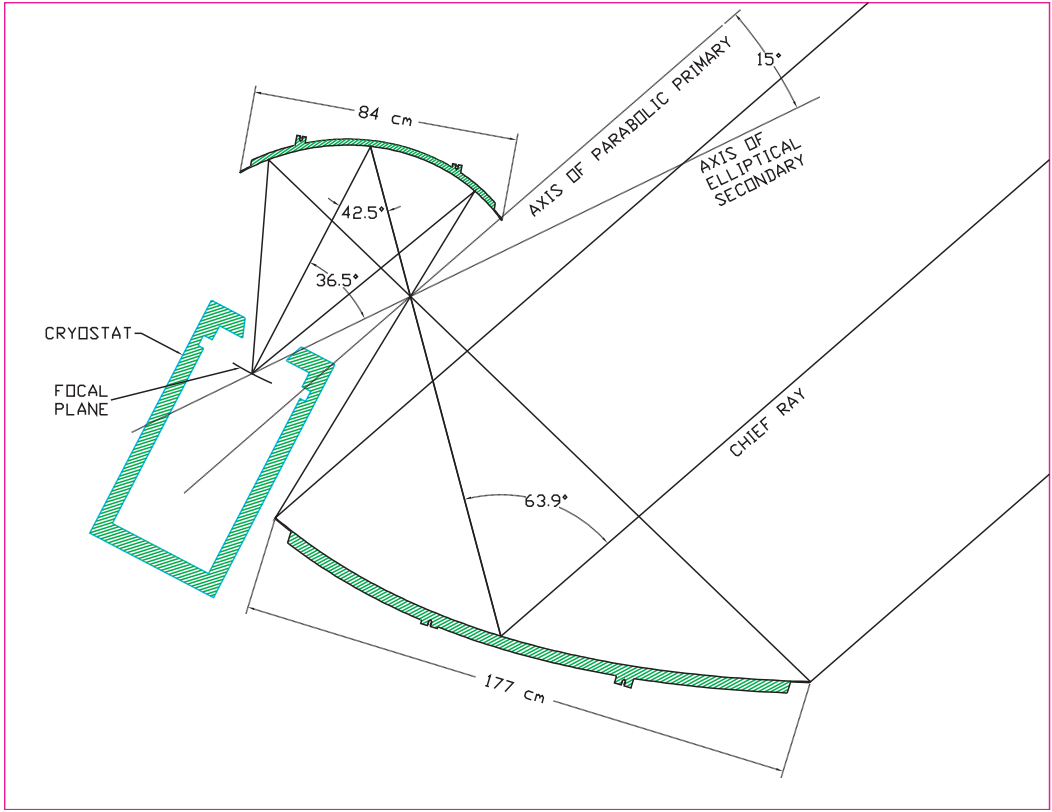}}
\caption{ A side view of the Archeops telescope and cryostat. The telescope
is a tilted Gregorian system consisting of a parabolic primary and elliptic
secondary. The primary axis of the ellipse is tilted with respect to the 
axis of the parabola such that the system satisfies the Dragone condition. 
The system provides an RMS wave-front error of less than 6\% at wavelengths
between 2 and 0.85 mm, for beam sizes between 8 and 5 arcminute, respectively. 
        }
\label{fig:telescope_schematics}
\end{figure}

\begin{figure}%[t]
\resizebox{\hsize}{!}{\includegraphics{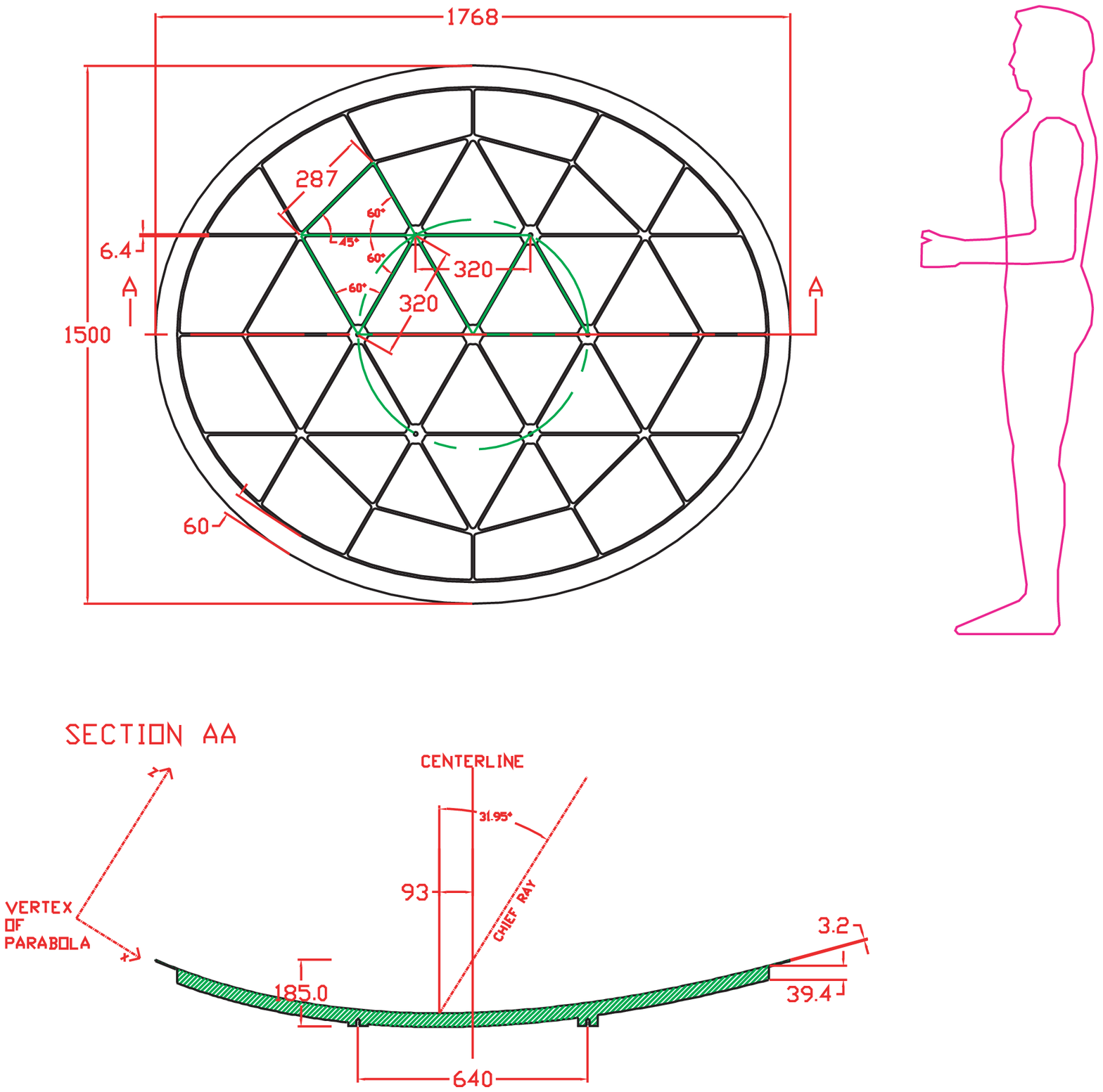}}
%\centerline{\epsfig{file=arch_primary_back_side.eps,width=4.5in,angle=0}}
\caption{ A side and back view of the Archeops telescope primary 
mirror.         }
\label{fig:primary}
\end{figure}

Table~\ref{tab:teles} lists the specifications of the telescope and a
schematic is shown in Figs.~\ref{fig:telescope_schematics} and
\ref{fig:primary}.  The primary mirror is the aperture stop of the
system and is defined as the intersection of a cylinder of 1.5~meter
diameter with an off-axis section of a paraboloid.  It has an
elliptical shape with a length of 1.8~meters and 1.5~meters for the
major and minor axes, respectively. The oversized secondary is close
to circular, with a diameter of about 0.8~m.

Both mirrors were milled from 8~inch thick billets of aluminum 6061-T6
and were thermally cycled twice during machining to relieve internal
stresses.  The 5~mm thick reflecting surfaces are supported by a
honeycomb rib structure that reduces deflections due to gravity and
milling machine pressure (during machining), and which ensures that
the material does not exceed its yield strength under the expected
load of $10 g$ during parachute shock at the end of the flight.  The
primary mirror construction is shown in Fig.~\ref{fig:primary}.  The
primary and secondary mirrors weigh 45~kg and 10~kg, respectively.  The
mirrors were hand polished to provide the required figure
accuracy. The local surface roughness is less than 2 \micron\
roughness average~\cite{Green:1996}, and the overall figure accuracy
is 50 \micron\ rms.
\begin{table}
\begin{center}
\begin{tabular}{l|l} \hline \hline
\multicolumn{1}{c|}{Parameter} & Value \\ \hline
f\#$^{a}$                       & 1.7  \\ 
Plate Scale$^{a}$              & 35 mm/deg \\ \hline
%Beam FWHM                 &    \\
%\hspace{0.5cm}                 143 GHz  &  8.0 arcmin \\ 
%\hspace{0.5cm}                 217 GHz  &  5.5 arcmin \\ 
%\hspace{0.5cm}                 353 GHz  &  5.0 arcmin \\ \hline
Primary Mirror            &    \\
\hspace{0.5cm}                 conic constant & -1 \\
\hspace{0.5cm}                 radius of curvature of paraboloid & 160 cm \\ 
\hspace{0.5cm}                 major axis of mirror & 177 cm \\
\hspace{0.5cm}                 minor axis of mirror & 150 cm  \\ 
\hspace{0.5cm}                 weight   & 45~kg \\ \hline
Secondary Mirror            &    \\
\hspace{0.5cm}                 conic constant & -0.18 \\
\hspace{0.5cm}                 radius of curvature   & 53 cm \\ 
\hspace{0.5cm}                 tilt of major axis$^a$ & 15 degrees \\
\hspace{0.5cm}                 major axis of ellipsoid & 130 cm \\
\hspace{0.5cm}                 axial magnification of elliposoid & 2.5 \\
\hspace{0.5cm}                 major axis of mirror & 84 cm \\
\hspace{0.5cm}                 minor axis or mirror & 79 cm \\ 
\hspace{0.5cm}                 weight   & 10~kg \\\hline
Image Surface              &  \\
\hspace{0.5cm}                 conic constant & 0 (= spherical) \\
\hspace{0.5cm}                 radius of curvature & 101 cm \\ \hline \hline
\multicolumn{2}{l}{$^{a}$ Value at the center of the focal plane} \\ \hline
\end{tabular}
\end{center}
\caption{Design and construction parameters for the Archeops
telescope. }
\label{tab:teles}
\end{table}

\subsection{Focal plane}\label{ss:FP}

%The focal plane configuration during the flight in Trapani is shown in 
%Fig.~\ref{fig:figfocal}. It consisted of a total of six photometers, 
%three at the 143~GHz band (B1-1,2,3 $\equiv$ Band 1; photometers 1,2,3), 
%two at the 217~GHz band (B2-4,5) and one at the 353~GHz band (B3-6).
The focal plane configuration during the flight in Trapani consisted
of a total of six photometric pixels, three at the 143~GHz band
(B1-1,2,3 $\equiv$ Band 1; pixels 1,2,3), two at the 217~GHz band
(B2-4,5) and one at the 353~GHz band (B3-6).  A photometric pixel is
defined as the assembly of one bolometer and its front cold optics
(see Fig.~\ref{fig:cold_optics}).  The angular separation between the
photometric pixels in the focal plane was roughly
50~arcmin. Fig.~\ref{fig:figfocal} shows an inflight measurement
of the focal plane geometry (see
section~\ref{sec:focal_plane_geometry_and_beams}). For Kiruna flights,
it is anticipated that the angular separation between nearby
bolometers will be 30~arcmin.
%\begin{figure}
%\resizebox{\hsize}{!}{\includegraphics[angle=0]{focalplane_nb.epsi}}
%\caption{The focal plane configuration during the Archeops flight
%in Trapani and the
%measured beam shapes. The white pluses show the expected location 
%of the photometers relative to the center one. The white crosses are 
%the centers of a Gaussian fit to the measured beams (preliminary results).  
%The measurement is discussed in section~\ref{sec:focal_plane_geometry_and_beams}
%}
%\label{fig:figfocal}
%\end{figure}

\subsubsection{Photometric pixel layout}

\begin{figure*}[t]
\resizebox{\hsize}{!}{\includegraphics[angle=270]{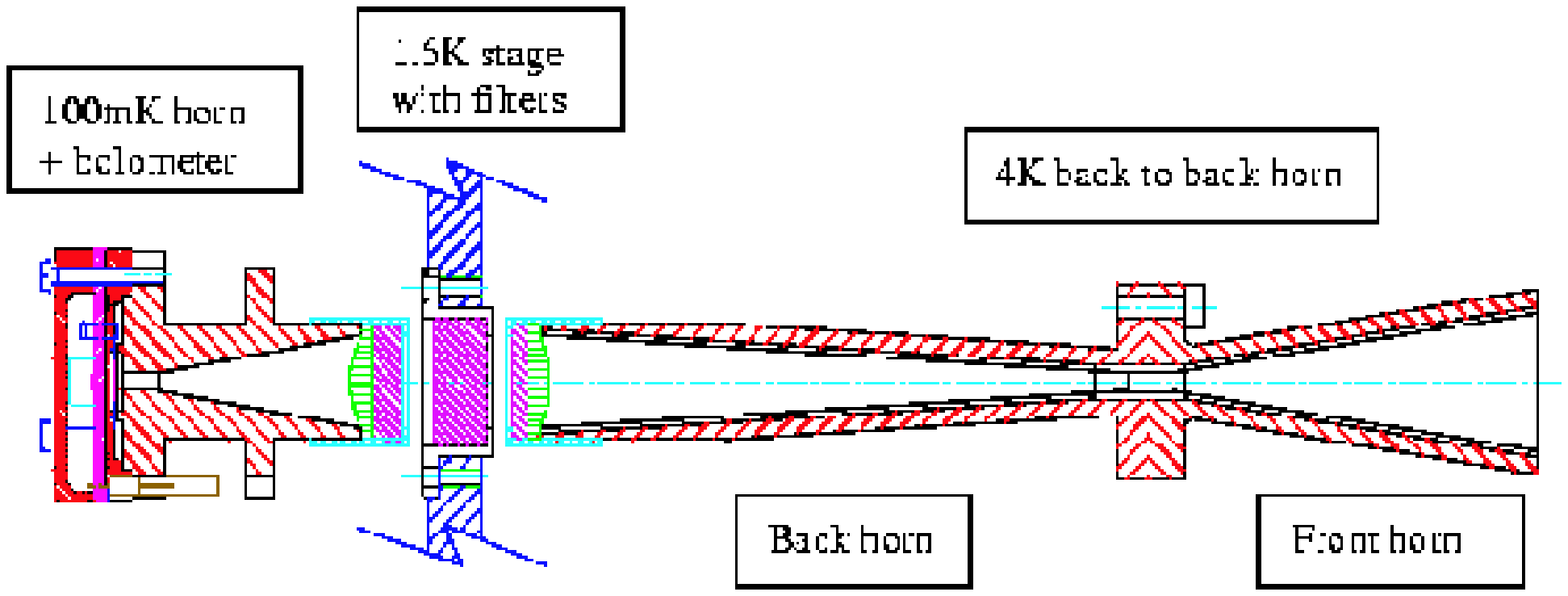}}
\caption{Optical configuration for a single photometric pixel}
\label{fig:cold_optics}
\end{figure*}

The configuration of the focal plane is similar to that being
developed for Planck~HFI.  For CMB anisotropy measurements, control of
spectral leaks and beam sidelobe response is critical. Archeops
channels have been specifically designed to maximize the sensitivity
to the desired signal, while rejecting out--of--band or out--of--beam
radiation.  Developing on a design put forward for a previous proposed
mission \cite{Church:1996}, we have chosen to use a triple horn
configuration for each photometric pixel, as shown schematically in
Fig.~\ref{fig:cold_optics}.  In this scheme, radiation from the
telescope is focussed into the entrance of a back--to--back horn pair.
With no optical components in the path, control of the beam is close
to ideal, as will be shown later.  A lens at the exit aperture of the
second horn creates a beam--waist where wavelength selective filters
can be placed.  Finally, a second lens on the front of the third horn
maintains beam control and focuses the radiation onto the spider
bolometer placed at the exit aperture.  A convenient aspect of this
arrangement is that the various components can be placed on different
temperature stages in order to create thermal breaks and to reduce the
level of background power falling onto the bolometer and fridge.

In Archeops, the back--to--back horn pair is located on a cold plate
cooled by Helium vapor.  In flight it reaches a temperature of about
10~K because of the optical load from the telescope and the
atmosphere. This results in some power loading on the detectors from
the horns.  A model of the photometric performance of the system shows
that even with this high horn temperature, the detector is limited by
the power radiated by the telescope and atmosphere, which determine
the fundamental limit for this sub--orbital experiment.  The
back--to--back horn pair is constructed from two horns separated by a
waveguide section.  Sidelobe response, beamwidth on the sky and
spillover are accurately controled by the design of the front horn.
The waveguide is critical in that it defines the spectral high--pass
cut--on for the band and controls the allowable modes that propagate
from the telescope to the detector. Low pass filters placed
at the exit of the horn pair reject all high frequency radiation from
the inner stages and bolometer.

To further shield the dilution refrigerator from unwanted thermal
power, it is surrounded by a shield at 1.6~K.  With our optics scheme
we were able to place more low pass filters at this point to further
reject unwanted radiation from the inner sanctum where the 100~mK
detectors are located.

\begin{figure}%[t]
\resizebox{\hsize}{!}{\includegraphics[angle=0]{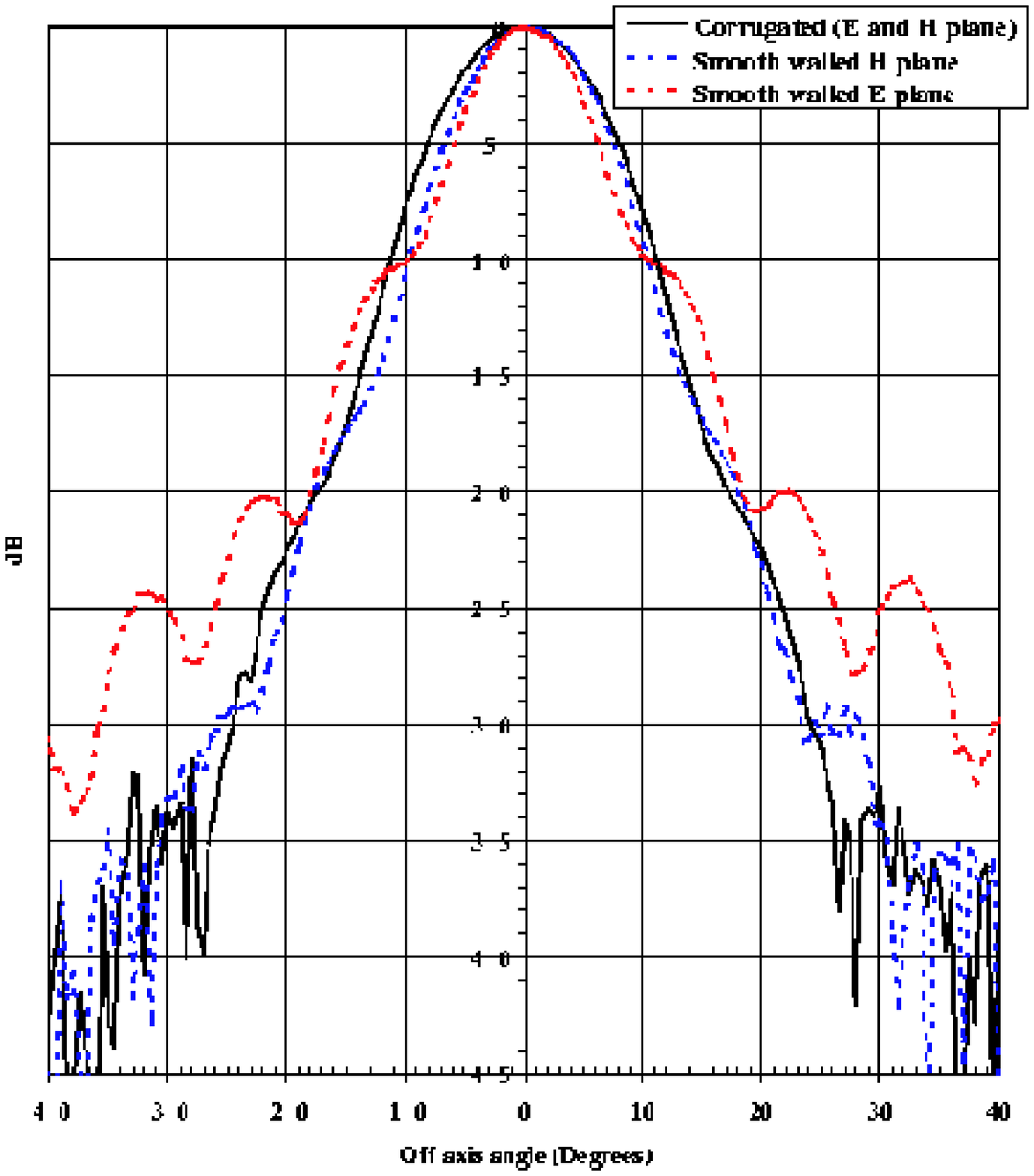}}
\caption{Comparison of beam patterns (E and H plane polarisation)
measured at 150~GHz for corrugated and smooth walled back to back
horns. This figure shows the asymmetry between the 2 polarisations in
the case of a smooth walled horn which will increase the sidelobe
level, while in the case of a corrugated horn, the 2 polarisations are
identical (only one is shown) down to the level of 35~dB
(measurement limit).}
\label{fig:antenna_profile}
\end{figure}

A second lens in front of the final 100~mK horn refocuses the beam,
which is then coupled to the bolometric detector by the horn and a
further waveguide section.  The diameter of this waveguide is set to
be larger than the first back--to--back waveguide to ensure that it does
not modify the mode propagation of the feed system.  The low pass band
edge-defining filter was also located on the front of this horn at 100
mK.

\begin{figure}%[t]
\resizebox{\hsize}{!}{\includegraphics[angle=0]{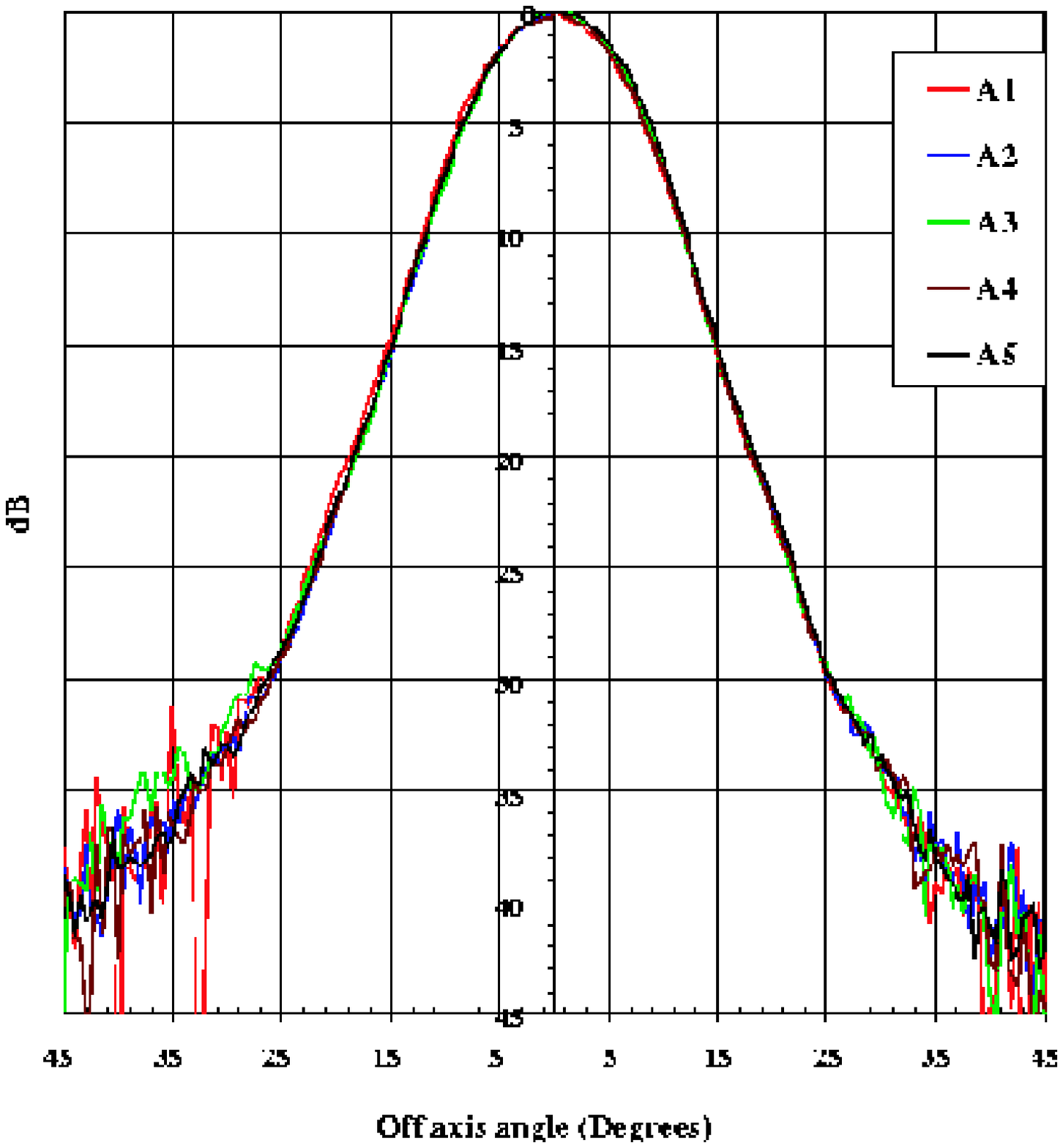}}
\caption{Measured beam patterns at 180~GHz for all the corrugated
horns used during the 1st flight of Archeops in Trapani. We can notice the good
mechanical reproducibility in the horn manufacture which gives
identical beams for all the horns.}
\label{fig:meas_antenna_profile}
\end{figure}

For the Trapani test flight, all the 143 and 217~GHz photometric
pixels used identical conical corrugated back-to-back horns, which
were single moded at 143 GHz but allowed a few modes to propagate at
217 GHz. The 353~GHz photometric pixel used a multi-moded smooth
walled back-to-back horn. Measurements and models have shown
\cite{Maffei:2000} that corrugation in the horns produce noticeable
improvement in the symmetry of the antennal profiles between the E and
H polarization patterns (see Fig.~\ref{fig:antenna_profile}), and in
the sidelobe levels.
The use of the smooth walled horn for 353~GHz was necessary because
the only corrugated horns available were those designed for 143~GHz,
which have poor transmission at these much higher frequencies
\cite{Colgan:2000}.  Proper single mode corrugated feeds will be
made for the Kiruna flights.  Fig.~\ref{fig:meas_antenna_profile}
shows the measured antenna response for the corrugated horns that were used
during the Trapani flight.  We estimate that we have achieved -25~dB telescope edge
taper with the telescope/horns combination. 
New profiled--flared horns that will be used in the
Kiruna flights will achieve -30~dB.

\subsubsection{Spectral Filtering}

\begin{figure}%[t]
%%% \resizebox{0.8\hsize}{!}{\includegraphics{tech_paper_arch_horn.eps}}
%%% \caption{Transmission of all filter components for the 143~GHz
%%% channel, (a: top) logarithmic scale, (b: bottom)  linear scale}
\resizebox{0.8\hsize}{!}{\includegraphics{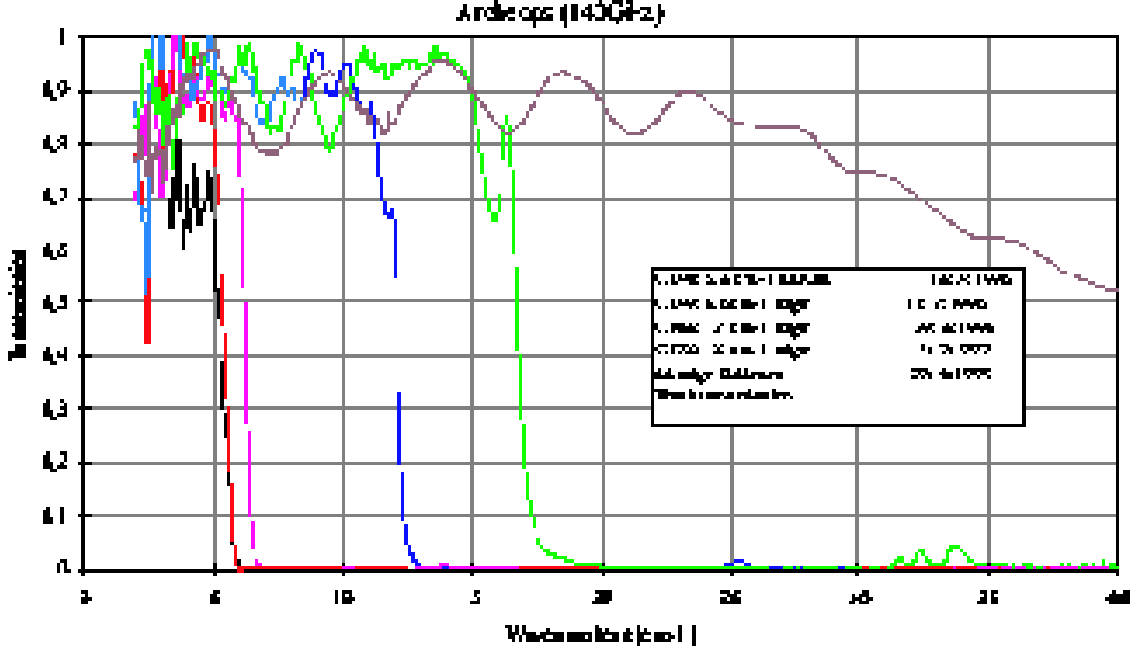}}
\caption{Transmission of all filter components for the 143~GHz
channel. Total transmission is in black}
\label{fig:A_spectral_band}
\end{figure}
\begin{figure}%[t]
\resizebox{\hsize}{!}{\includegraphics{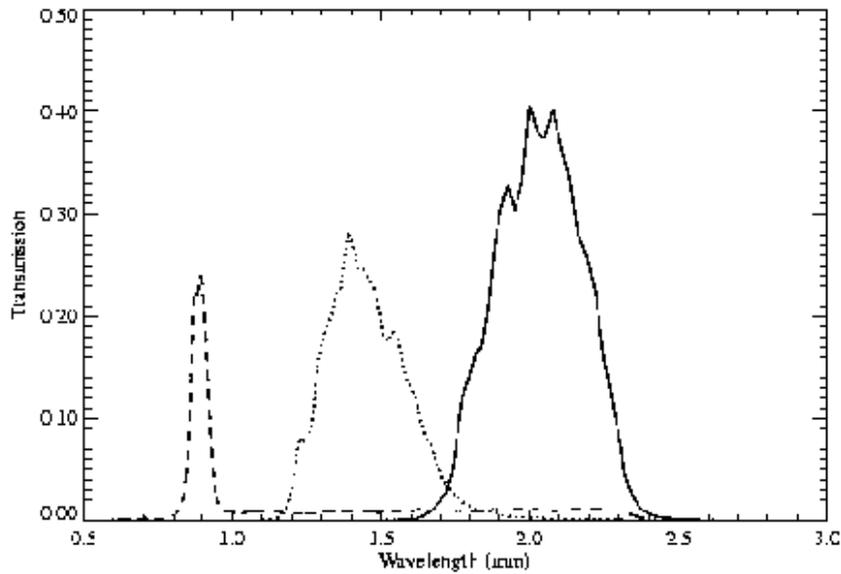}}
\caption{Total transmission of Archeops channels obtained by combining
separate measurements of each filtering components (10~K, 1.6~K,
0.1~K). Channel 1 (nominally taken, for calibration purpose, at
143~GHz) is centered at 150~GHz roughly covering 134 to 164~GHz at
half maximum so that the bandpass is $\Delta\nu/\nu=0.20$.  Channel 2
(nominally at 217~GHz) is centered at 211~GHz roughly covering 188 to
236~GHz at half maximum so that the bandpass is
$\Delta\nu/\nu=0.23$. Channel 3 (nominally at 353~GHz) is centered at
341~GHz roughly covering 329 to 354~GHz at half maximum so that the
bandpass is $\Delta\nu/\nu=0.074$. Due to lack of time before launch,
this latter bandpass could not be optimised. The measurements are not
accurate for wavelengths larger than about 2.5~mm.}
\label{fig:Arch_band}
\end{figure}

To obtain a well-defined band edge and ensure high rejection for
UV-optical-IR wavelengths, a series of metal mesh low--pass
interference filters were used.  Each filter was manufactured from
several layers of photolithographically etched copper-patterns on
polypropylene substrates and then hot pressed together to form a
rugged filter \cite{Lee:1996}.  The back--to--back horn waveguide
defines the low frequency cut-on, and hence by choice of filters, the
number of waveguide modes propagated.  Fig.~\ref{fig:A_spectral_band}
shows the measured response on a linear plot for all the filter
components used in the 143 GHz channel. Out--of--band rejection is
better by a factor $10^{-10}$ in the submillimetre domain.  Tests on
similar filter stacks for Planck HFI show that the good rejection is
maintained through the near IR, optical and UV regions.  It is clear
that the total in--band spectral transmission is typically 55\%. The
complete expected transmission of the cold optics filtering scheme is
shown in Fig.~\ref{fig:Arch_band}.

\subsubsection{Bolometric detectors}

The spider web bolometric detectors were developed by JPL/Caltech as
part of the Planck HFI development.  These bolometers have high
responsivity, low NEP, fast speed of response and a low cross section
to cosmic ray hits. The bolometers are made by micro-machining silicon
nitride to leave a self-supported, metallized spider web mesh absorber.
A neutron--transmutation--doped Ge:Ga thermistor is then indium bump
bonded at the web center to sense the temperature rise which results
from the absorption of IR power \cite{Bock:1996}, \cite{Mauskopf:1997}.
The specific requirement for the Archeops detectors, which are cooled
to 100~mK, is that the sensitivity should meet with the expected
background limited photon noise $NEP_{BLIP}=1.5\times
10^{-17}\zu W/Hz^{-1/2}$ and have a response time of 1.6~ms, dictated
by the instrument spin rate.

\begin{figure*}
\resizebox{\hsize}{!}{\includegraphics[angle=0]{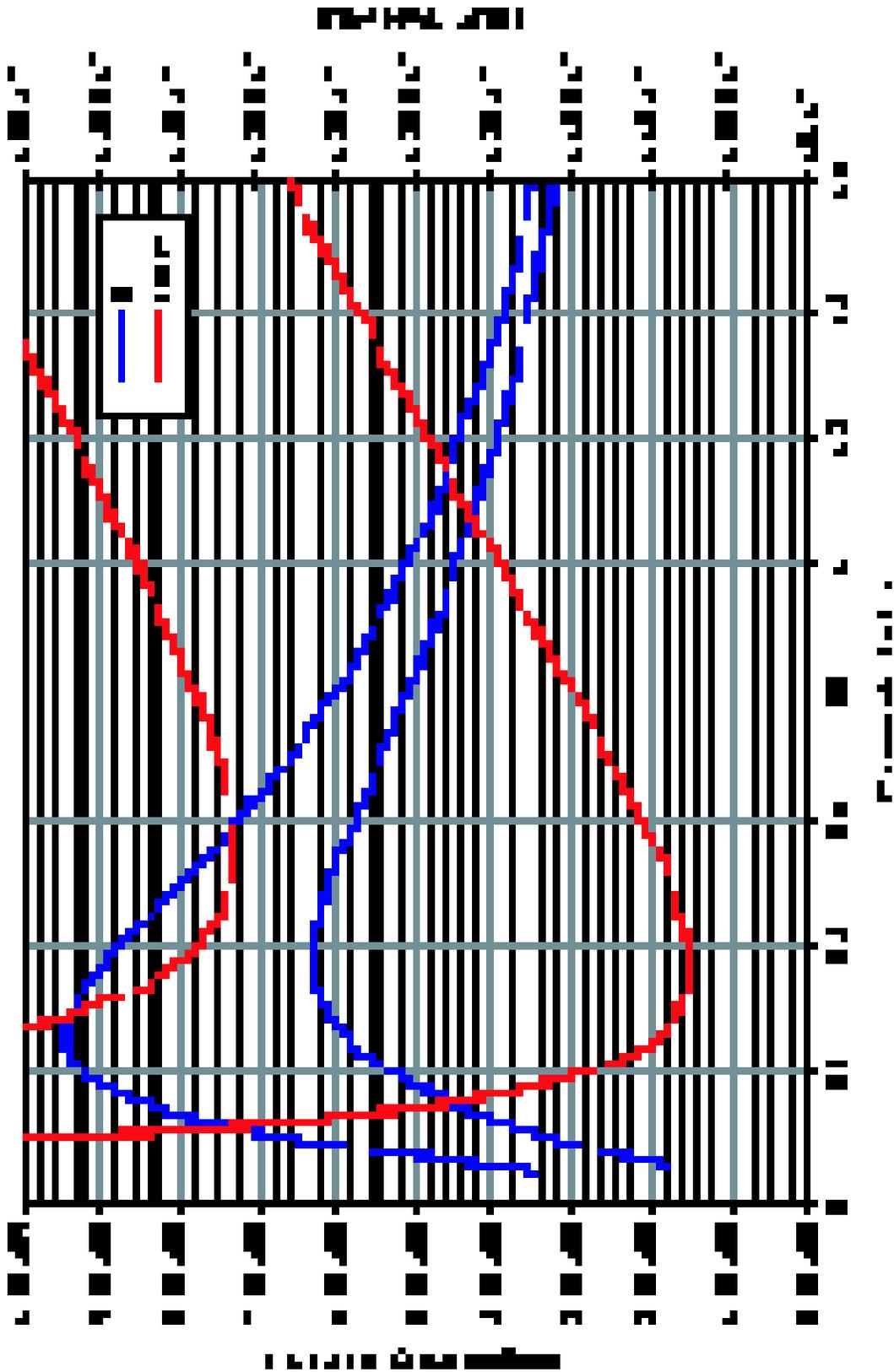}}
\caption{The $NEP$ and Responsivity versus bias current for
pixels B1-1 and B3-6 at a base temperature of 98~mK. The bolometers are
optically loaded with radiation from a 300~K blackbody attenuated to
levels expected in flight. The $NEP$ includes Johnson and phonon noise
contributions. The optimum bias current, $I_{opt}$, is when the $NEP$
is smallest.}\label{fig:bolo}
\end{figure*}

\begin{table}
\caption[]{Summary of the measured optimum bias current $I_{opt}$, and
the values of $NEP$, Responsivity ($Resp$) and optical time response
($\tau$) of some bolometers operating at this bias current and a base
temperature of 98~mK. The numbers are accurate to about 10~\%.}
\begin{tabular}{cllll}
Bolometer &   $I_{opt}$ & $10^{17} \times NEP$ &   $Resp$ & $\tau$ \\
          &   (nA)      & ($\zu W Hz^{-1/2}$) & (MV/W)   & (ms) \\
\hline
B1-1  & 0.35 & 1.07  & 920 & 3.5\\
B1-3  & 0.32 & 1.10  & 920 & 1.0\\
B3-6  & 0.55 & 1.40  & 630 & 3.0\\
\hline
\end{tabular}
\label{tab:bolo}
\end{table}

In preparation for the Trapani flight, three of the bolometers
(photometric pixels~B1-1, B1-3 and B3-6) were characterised as a
function of operating temperature around 100~mK using an adiabatic
demagnetization refrigeration (ADR). Photometric pixel B1-1 was
designed and used in Band~1, 143~GHz.  Bolometer 3-6 was designed for
use in Band~2, 217~GHz , but was used in Band~3 at
353~GHz. Photometric pixel 3 uses a prototype Planck HFI 143~GHz
bolometer which was used in Band~1, 143~GHz. These three detectors
were characterised optically in the ADR test-bed using prototype horns
and filters optimised for 143~GHz operation \cite{Woodcraft:2000} and
\cite{Sudiwala:2000a}. The $NEP$ and responsivity versus bias current
for photometric pixels B1-1 and B3-6 operating at a base temperature
of 98~mK and under flight simulated optical loading conditions are
shown in Fig.~\ref{fig:bolo}. The superimposed model fit curves have
been calculated from the measured bolometer parameters using methods
described by \cite{Sudiwala:2000b}. These data, together with the
measured optical response times, are summarised in
Tab.~\ref{tab:bolo}.

\subsection{Cryostat}

\begin{figure}
\resizebox{\hsize}{!}{\includegraphics{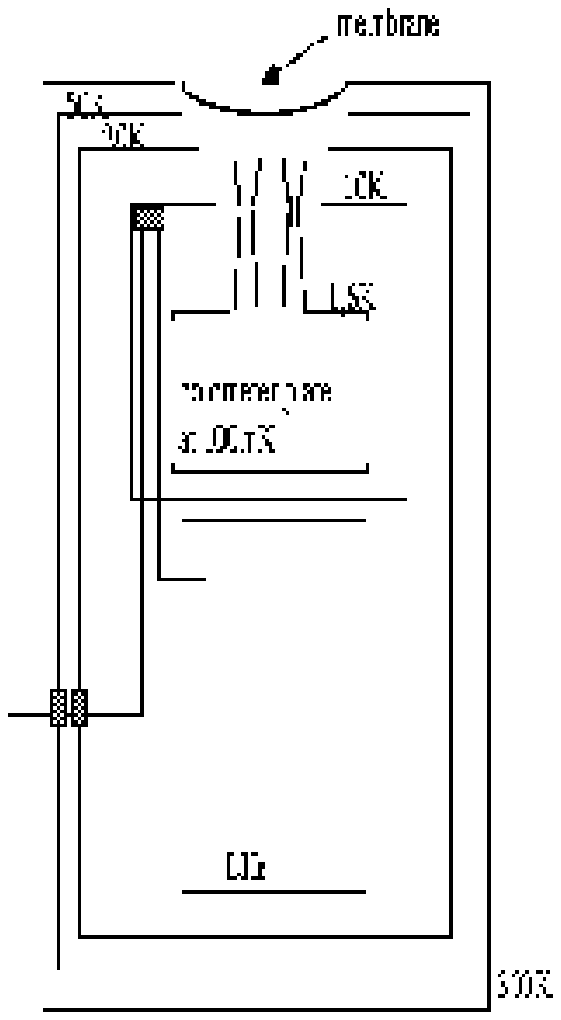}}
\caption{Schematic drawing of the cryostat: the membrane opens the
cryostat towards the secondary mirror.}
\label{fig:cryostat1}
\end{figure}

\begin{figure}
\resizebox{\hsize}{!}{\includegraphics[angle=90]{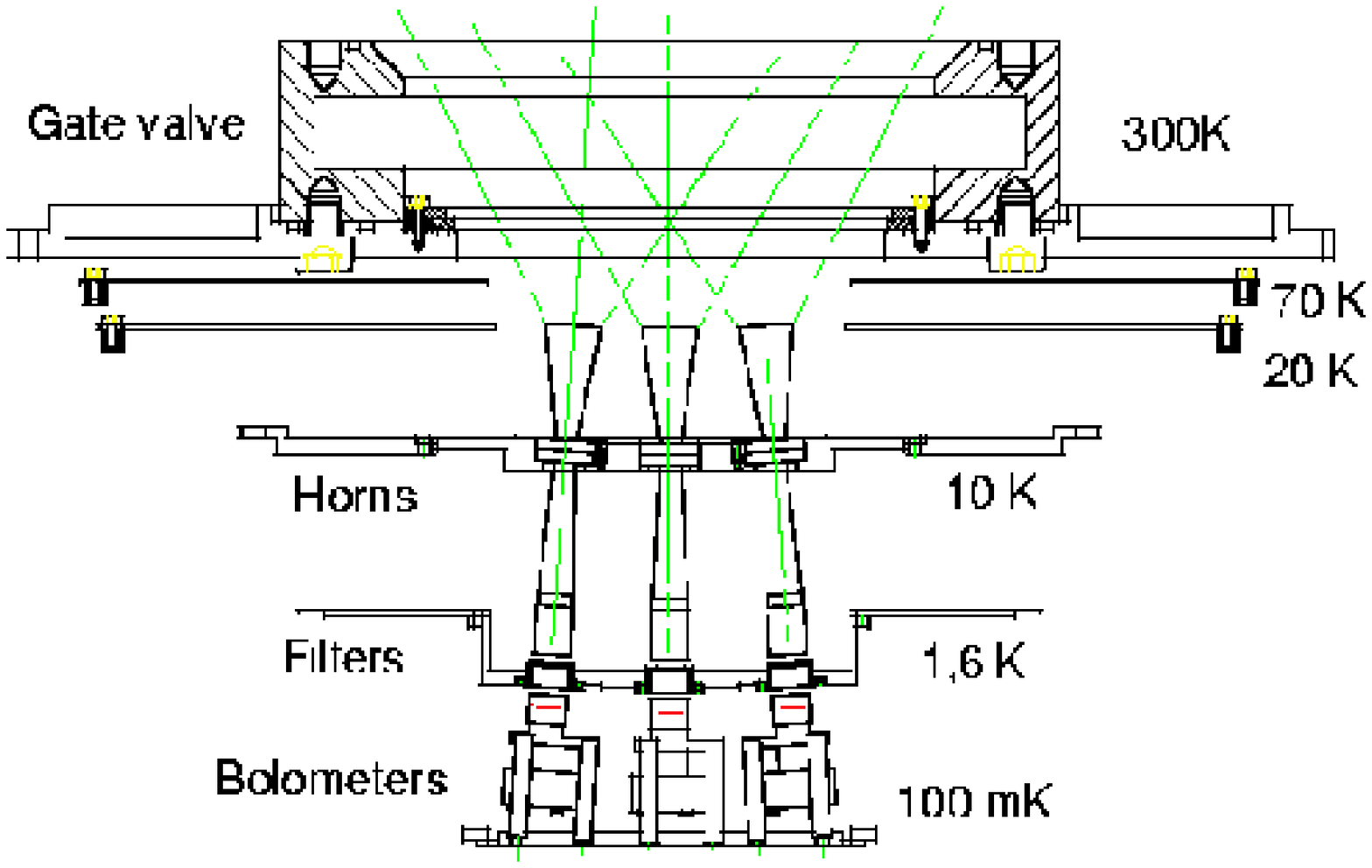}}
\caption{Cryogenic and mechanical implementation of the cold optics in
the cryostat.}
\label{fig:cryostat2}
\end{figure}

%I / 3 / 1  Cryostat description
The focal plane is cooled to 100~mK by means of an open cycle dilution
refrigerator designed at the CRTBT (Centre de Recherche sur les Tr\`es
Basses Temp\'eratures) in Grenoble (France) by \cite{Benoit:1994a}
(see also~\cite{Benoit:1994b}).  The dilution stage is placed in a low
temperature box placed on a liquid Helium reservoir at 4.2~K. The top
part of this box contains the entrance horns and receives a
significant amount of heat power from near infrared radiation ($300 -
700\;$ mW).  We therefore put in place a heat exchanger using exhaust
vapours from the helium tank to maintain the horns near 10K
(Fig.~\ref{fig:cryostat1} and \ref{fig:cryostat2}). The entrance is
protected from radiation by two vapour cooled screens with openings
for the input beam. The filters are placed on the horns at 10 K, on
the 1.6 K stage (cooled by Joule--Thompson expansion of the dilution
mixture) and on the 100 mK stage. The temperatures of each stage are
monitored with thermometers: carbon resistance and NbSi metal
insulation transition thermometers (see Sect.~\ref{subs:th}).

The bolometers are placed on the 100~mK stage supported by Kevlar
cords. The dilution fluids arrive through two small capillary tubes,
after mixing, and cool the stage using a small heat exchanger, finally
exiting through a third capillary. These three capillary tubes, two
extra capillary tubes used for precooling and the electric wires (9
shielded cables with 12 conductors each) are soldered together,
forming the continuous heat exchanger disposed around the 100 mK
stage. The Planck HFI instrument design is similar.  Input flow is
controlled by an electronic flow regulator; the output mixture is
pumped with a charcoal pump placed inside the liquid helium (1 liter
box filled with charcoal). An electronic regulator is used to maintain
constant pressure at one atmosphere in the helium tank.

% I / 3 / 2  Thermal fluctuation control
\subsubsection{Thermal fluctuation control}

\begin{figure}
\resizebox{\hsize}{!}{\includegraphics{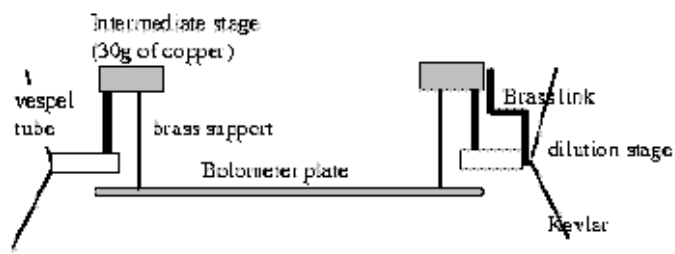}}
\caption{Schematic drawing of the thermal filtering of the 100~mK stage .}
\label{fig:th_filtering}
\end{figure}

At the junction of the $^4$He and $^3$He tubes, a random sequence of
concentrated and diluted $^3$He phases is created, causing a
temperature fluctuation of about 100 $\mu$K and producing noise in the
bolometers.  The bolometer plate is connected to the 100~mK dilution
stage through a link with controlled thermal conduction
(Fig.~\ref{fig:th_filtering}). An intermediate stage along this link
can be used to regulate the temperature.

The first thermal link is composed of a vespel support and a brass
link in parallel in order to regulate thermal conduction. Combined
with a copper mass, the time constant of the intermediate stage is
about 1 second. The main filtering is due to the bolometer plate
specific heat itself ($\sim 10^{-3}\;$ J/K) . The time constant at
this point is around 30 seconds.  To obtain the best possible
stability, we use a regulation on the intermediate stage (to suppress
low frequency fluctuations) and passive filtering between the
intermediate stage and the bolometer plate to suppress high frequency
fluctuations.  In fact, during the test flight, the dilution stage did
not reach the base temperature, but stopped cooling at 112~mK (because
of a small leak in the cryostat).  The regulation stage therefore
remained inactive and we only used passive filtering.

%I / 3 / 3  Thermometers
\subsubsection{Thermometers}\label{subs:th}

Anderson insulators were used to measure the temperature at
100~mK. They were made at CSNSM Orsay \cite{Dumoulin:1993}.  These
are thin layers of Niobium--Silicium deposited on saphire substrate
and are very sensitive at 100~mK, where their impedance reaches a few
M$\Omega$.  The read--out electronics are the same as for the
bolometers.

\subsection{Stellar Sensor}\label{subs:ss}

A custom star sensor has been developed for pointing reconstruction in
order to be fast enough to work on a payload rotating at 2--3 rpm.  At
this spin rate, the use of a pointed platform for the star sensor is
impractical. Each independent beam (8~arcmin wide) is scanned by the
mm--wave telescope in about 10~ms, establishing a detector response
time that excludes the use of present large--format CCDs. We decided
therefore to develop a simple night sensor, based on a telescope with
photodiodes along the boresight of the mm--wave telescope.  Thus, like
the millimetre telescope, the star sensor scans the sky along a circle
at an elevation of $41^o\pm 1^o$.

A linear array of 46 sensitive photodiodes (Hamamatsu S-4111-46Q) were
placed in the focal plane of a 40~cm diameter, 1.8~m focal length
parabolic optical mirror.  Each photodiode has a sensitive area of 4~mm (in
the scan direction) by 1 mm (pitch in the cross--scan direction).  The
dominant aberration of this simple telescope is coma, which is
contained within a 0.2~mm diameter spot for the photodiodes at the
edges of the array.  The line of photodiodes is perpendicular to the
scan (and so is aligned along elevation), and covers 1.4~degrees on
the sky, with about 7.6~arcminutes (along the scan) by 1.9~arcminutes
(cross--scan) per photodiode.  A top baffle, painted black inside and
located above all nearby payload structures, prevents stray radiation
(direct and reflected from the gondola) from entering the telescope
tube. A small baffle near the detector window further prevents stray
radiation from reaching the photodiodes.

During one rotation of the payload, the star sensor scans a full
circle, 1.4$^o$ wide at 41$^o$ elevation.  On average, this circle
contains about 50 to 100 stars with magnitude brighter than 7.  The
limiting magnitude of the sensor depends on the integration time
available, as well as on the optical filters placed in front of the
array (to reject parasitic signals) and on the electronic read--out
noise.  For night--time operations of the test flight, we decided to
avoid the use of optical filters in front of the sensor. The
photodiodes are used in a photovoltaic configuration for minimal
noise. The signal from the photodiodes is AC coupled (1.06~Hz cut--in
frequency) to remove large--scale brightness gradients, and amplified
by an low noise operational amplifier (OPA129P) connected as a
current--to--voltage converter.  The amplifier features a 1 G$\Omega$
feedback resistor with a 0.5~pF capacitor in parallel to eliminate
gain peaking.  A second amplification stage (G=56) and a low--pass
filter (second order, 33.1~Hz cut--off) are applyed to cover the
dynamical range of a 12 bit A/D converter.  The data are sampled 180
times per second and compressed (with loss) to 4 bits for each
photodiode. In this configuration, the sensor can detect stars of
magnitude 6-7 in 5~ms of integration.

\begin{figure}
\resizebox{\hsize}{!}{\includegraphics{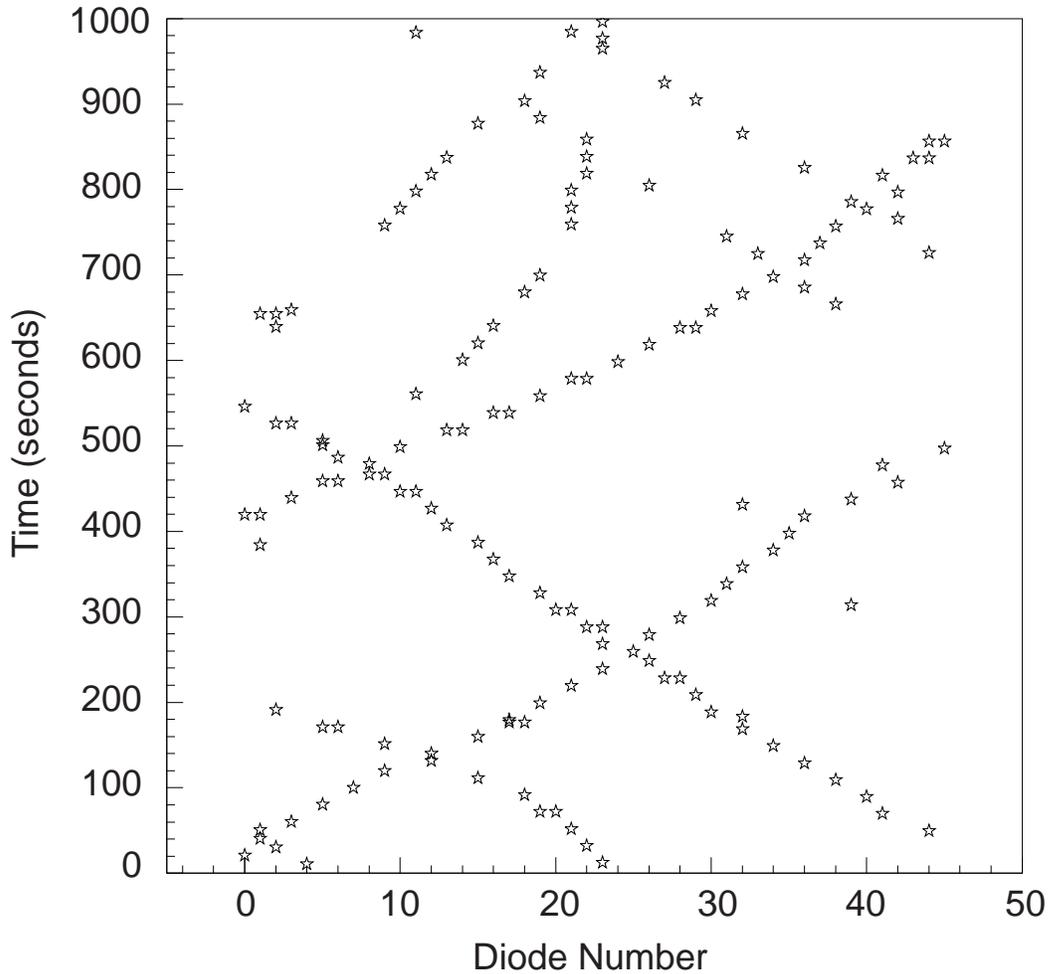}}
\caption{This graph shows the diode number versus time of large 
amplitude candidates as seen by the star sensor.
Bright stars trigger the star detector at every pass (20 second intervals).
They are seen as "tracks" in this plot: as the sky moves in front of the
balloon, star triggers move from one photodiode to the next.}
\label{fig:fss}
\end{figure}

The goal of the star detector software is to extract, from the
time--sampled photodiode current values, candidate stars with
detection time, measured flux, and quality criteria.  This software
should be able to run continuously during flight. For this we
implemented an algorithm that treats independent photodiode data
streams. The schematic of the algorithm is the following: for a given
photodiode, when the program starts, it fills an array with the first
48 raw data samples and copies them to an ``Analyzed Array'' (AA).
Among these, the first 32 samples form the ``prepulse''.  The samples
from 36 to 44 are named the ``pulse array''.  Over the prepulse, we
compute an average and a slope. We subtract these contributions from
the Analysed Array, and then compute the root--mean--square noise over
the prepulse (RMS).  We search the pulse array for two consecutive
samples larger than 6 times the RMS.  If found, we look for the pulse
maximum. We then compute the full--width at half--maximum, fit a
parabola over the maximum and store the fitted time and amplitude. We
may also compute, from the hardware description and rotation speed of
the balloon, the expected pulse shape of an actual star event.  We
then fit this pulse shape to the data, and store a fitted amplitude,
time and $\chi^2$ for the candidate star. We add this data to the
list of time--ordered star candidates and make it available for the
second step of the software -- reconstruction of the telescope
pointing.  We iterate this algorithm over 46 photodiode data streams
and then push the data streams by 8 samples forward.  Selecting for
clarity the largest amplitude star pulses, we plot in Fig.~\ref{fig:fss} the
photodiode number of candidate star triggers versus time.  Bright
stars trigger the photodiodes at each rotation of the balloon over the
sky.  They drift over the photodiode array as the sky drifts in front
of the balloon. From the trigger rate we deduce that the stellar
sensor's limiting sensitivity is 6.6 magnitudes.

\subsection{Gondola} 

\begin{figure*}[t]
\resizebox{\hsize}{!}{\includegraphics[angle=90]{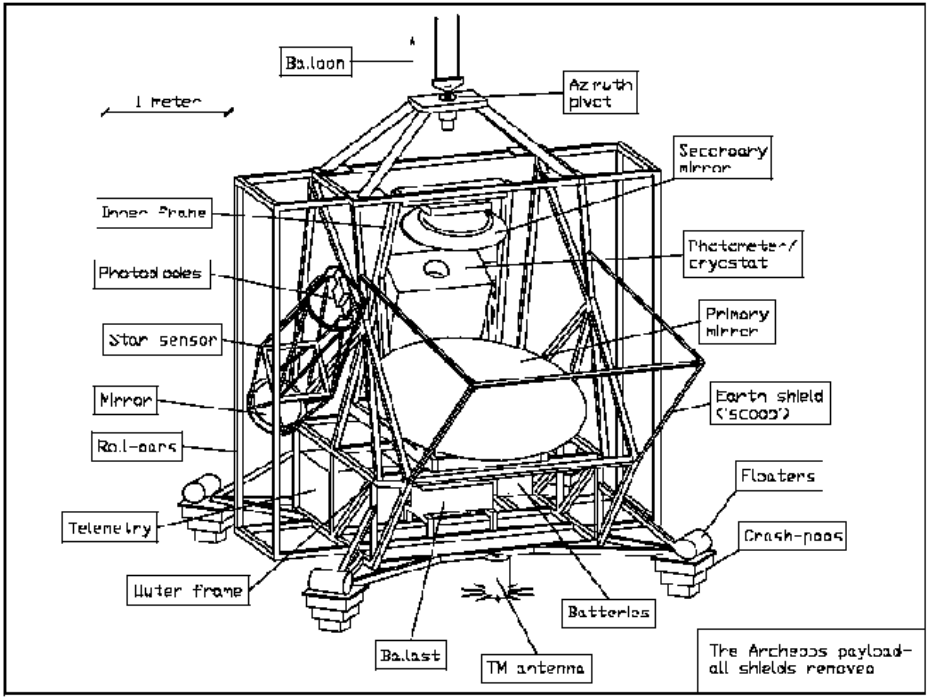}}
\caption{Schematic functional drawing of the gondola}
\label{fig:gondola}
\end{figure*}

The gondola is composed of two frames (see Fig.~\ref{fig:gondola}).  A
stiff inner frame is needed to support the two mirrors and the
photometer with negligible displacement under gravity, and subject
only to self--similar scaling under temperature changes.  It consists
of welded 6061T aluminum C profiles (same material as the mirrors).
The outer frame is lighter (built using aluminum square tubes -
50x50x2 mm) and connects the inner frame to the flight chain and
supports the star sensor and navigation hardware (telemetry, ballast,
batteries).  The mass of the scientific part of the payload is 500~kg;
the total Trapani test flight payload mass at launch was 1000~kg,
including 250~kg of lead grain ballast. The inertial moment around the
vertical axis is $I_{zz} \sim 1073\;$kg.m$^2$. The off-diagonal
elements of the inertia matrix are $\sim$ 50 times smaller than
$I_{zz}$. $I_{xx}$ and $I_{yy}$ are $881$ and $728\;$kg.m$^2$,
respectively.  We thus expect pendulation resonances at $\sim 1$~Hz,
significantly lower than the frequencies ($ \simgt 4-6\;$Hz) where the
acoustic peaks of the CMB anisotropy are encoded by the 2--3~rpm
azimuth scan.

The purpose of the Attitude Control System (ACS) is to spin in azimuth
the entire payload at 2--3~rpm.  To achieve this, we use an azimuth
pivot mounted on top of the gondola. The pivot connects the flight
chain of the balloon to the payload through a thrust bearing,
providing the necessary degree of freedom for payload spin. Two deep
groove bearings provide stiffness against transverse loads to the
rotating steel shaft inside the pivot.  A 60~mm diameter thrust
bearing ensures little dynamical friction, of the order of 0.5~Nm for
a 2 rpm rotation and 1 ton of suspended mass.  The pivot includes a
torque motor that acts against the flight chain to spin the payload.
After initial acceleration, the motor provides just enough torque to
compensate the friction in the thrust bearing and the small residual
air friction.  We use a motor with a torque constant of 0.24 Nm/A.
With a payload of 1 ton suspended below the pivot, this is enough to
reach the nominal rotation speed in a few minutes, and to maintain it
at 3 rpm during the flight with an average current of 2 A.  The flight
chain connecting the payload to the parachute and to the balloon is a
long ($>$ 10 m) ladder made of two steel cables (separated by 0.5 m)
and rigid steps every 3 meters.  When loaded with a 1 ton payload,
this ladder acts as a spring with reaction constant $>$ 50 Nm/rad.
This is stiff enough to apply all the needed torque without risks of
twisting the ladder.  The pivot is strong enough to withstand shocks
of up to 10 g at parachute opening (with a 1 ton payload).

The rotation of the payload is monitored by a set of three orthogonal
vibrating structure rate gyroscopes that can detect angular speeds as
low as 0.1\deg/s.  These are sampled at 171~Hz by a 16 bit ADC.  A PID
feedback loop control is implemented in software, comparing the target
azimuth velocity to the measured velocity.  A phase modulated signal
(PWM) to drive the pivot motor is synthesized from the error signal to
drive the torque motor in the pivot.  The PWM signal (8 kHz) drives an
H bridge of high power MOSfets. This powers the azimuth motor with
high current (up to 7~A for tests and spin--up), while maintaining
very low internal power dissipation.

\subsection{Electronics}

The following adjustable parameters are used to control the
experimental setup; they correspond to commands that can be sent to
the payload:
\begin{description}
\item[-] modulation control for the 24 channels (bolometers and thermometers)
\item[-] focal plane temperature regulation heater
\item[-] rotation velocity for the gondola
\item[-] the main gate valve
\item[-] 6 heaters on the photometer box
\end{description}
The list of measurements:
\begin{description}
\item[-] 24 channels for bolometer signals (sampled at 171~Hz)
\item[-] 8 channels for temperature measurements on the focal plane
(sampled at 171~Hz)
\item[-] 4 channels for measuring the temperatures of the different
screens (sampled at 171~Hz)
\item[-] 3 gyroscopes (sampled at 171~Hz)
\item[-] 46 stellar sensor diodes (sampled at 171~Hz)
\item[-] GPS data (time, position and altitude) (sampled at 0.5~Hz)
\item[-] 8 flow meters and pressure gauges for fluid control (sampled
at 0.1~Hz)
\item[-] 6 thermometers inside the photometer box (sampled at 0.1~Hz)
\item[-] 8 thermometers for Helium level control (sampled at 0.1~Hz)
\end{description}

\subsubsection{Readout electronics}

The bolometers are biased using AC square waves by a capacitive
current source. Their output is measured with a differential
preamplifier and digitized before demodulation.  We use the boxes
already designed in preparation for Planck.  Each box can manage 6
bolometers.  During the Trapani test flight, we used 4 boxes for a
total of 24 channels:
\begin{description}
\item[-] 4 channels modified to measure the temperature of the
           different screens and to drive the heater currents.
\item[-] 20 channels using a cold JFET preamplifier:
\begin{description}
\item[-] 9 channels for temperature measurements on the focal plane
\item[-] 6 bolometer channels
\item[-] 5 unused channels
\end{description}
\end{description}
All modulations are synchronous and driven by the same clock. This
clock was also used for data readout, which was simultaneous for all
bolometers and thermometers. Modulation parameters can be controled by
telecommand.  Sampling of the raw signal is at 6.51 kHz before
demodulation. Demodulation was performed by the EPLD and sampled twice
per modulation period. We used a frequency of 85.7 Hz for modulation
and 171.3 Hz for sampling.

\subsubsection{On--board computer}

The on--board computer consists of 3 cards:
\begin{description}
\item[-] The main card (TCR) contains: the transputer T805 and an EPLD
Altera 9400 to control communications and to directly drive the 6
readout boxes; the voltage regulators for the power supply; and the
GPS module.
\item[-] the dilution card (DIL) contains: three rapid converters for
the gyros, a phase modulated signal (PWM) to drive the pivot motor,
one converter with a multiplexer for general control and a solid state
relay to drive electro--valves and heaters.  An EPLD Altera 9400
controls the card.
\item[-] the stellar sensor card (SST) contains: the 48 converters for
stellar sensor data acquisition.  An EPLD Altera 9400 controls the
acquisition.
\end{description}

\subsubsection{On--board data storage}

After compression, the data are written to a storage module.  The
nominal option for this module is a 2 Gbyte Flash Eprom memory made of
256 circuits of 8 Mbytes each (we used only 1~GByte for the Trapani
flight). A dedicated microprocessor is used to write the Eprom. The
data storage module is installed in a sealed box, pressurized at 1
atmosphere. The data are read after retrieval of the balloon.

%%% \begin{table*}
%%% \caption[]{Batteries}
%%% \begin{tabular}{ccccl}
%%% No. of batteries  &   Voltage (V) & Voltage after regulation &        Current
%%% (A)  &   Purpose \\
%%% \hline
%%% 2 * 3 &  9         &   5       &  1.5  &    on-board computer \\
%%% 2 * 3 &  9         &   5       &  1.6  &    bolometer readout \\
%%% 2 * 6 &  $\pm 18$  & $\pm 15$  &  0.5  &    bolometer readout \\
%%% 2 * 6 &  $\pm 18$  & $\pm 15$  &  0.5  &    dilution \& stellar sensor \\
%%% 2 * 6 &  $\pm 18$  &           &  0/1  &    heaters and valves \\
%%% \hline
%%% \end{tabular}
%%% \label{tab:batt}
%%% \end{table*}

\subsubsection{Batteries}
The power supply consists of 48 lithium batteries of 3~V and 36 Ah.
This corresponds to a total power from 63 to 100~W.  
%%% A description of
%%% the batteries is given in Table~\ref{tab:batt}. 
The theoretical
duration of the batteries is longer than 45~hours for each part.  The
batteries are mounted together in an insulated box with a heater to
maintain the temperature above -10$^o$~C.  A switching system allows
us to save the system in the event of a battery failure.

\subsection{Data Aquisition}

%I.5.1 - Data flow between the experiment and the transputer (link1)
\subsubsection{Data flow between the experiment and the transputer
(link1)}

The sampling frequency may be adjusted from 100~Hz to 200~Hz and
corresponds to a period $P_{\rm s}$ of 5 to 10~ms (corresponding to
half a modulation period). The number of digitisations, $N_{\rm d}$,
over half the modulation period can be adjusted from 36 to 48,
yielding a digitisation period $P_{\rm d}$ of between 100 and
280~$\mu$s.  At this rate the EPLD sends data to the transputer as 64
bit words, corresponding to a data flow of 230 to 640~kbit/sec.  

%%% The
%%% 64~bit format is as follows:
%%% \begin{tabular}{ll}
%%% 16 bits :  &     raw data from one bolometer \\
%%% 16 bits :  &     star sensor data\\
%%% 21 bits :  &     bolometer measurement after demodulation\\
%%% 2 bits  :  &     synchronisation\\
%%% 1 bit   :  &     status of  all static parameters \\
%%%            &     (valves, heaters ...)\\
%%% 8 bits  :  &     gyroscope values, pressures, temperatures, \\
%%%            &     GPS data\\
%%% \end{tabular}

\begin{table*}
\caption[]{Data block description}
\begin{tabular}{ccccl}
block &  block size & & & \\
type  &  Trapani & emitted blocks & flow rate &  information in block \\
      &  bytes   & \%             & kbit/sec  &   \\
\hline
1     &  136  &   10     & 0.258      & On--board log\\
2     &  276  &   7      & 0.366 & acq. status: clock, gain, modul.,\ldots\\
3     &  260  &   2       & 0.098 & dilution status\\
4     &  96   &   10    & 0.182       & GPS character string\\
5     &  4240 &   2      & 1.6078 & 1-period samples: raw bolom.
data\\
9     &  880   &  100 & 16.684    & gyroscopes\\
12 & 1384 & 100 & 26.240 & comp. bolom. (32 chan.*7~bit)\\ 
14 & 2404 & 100 & 8.6784 & comp. sensor data (4~bit/pixel)\\
\hline
\end{tabular}
\label{tab:data_fmt}
\end{table*}

\noindent Only the raw data of a single bolometer is transmitted at a
time; after each full modulation period ($P_{\rm m} = 2 P_{\rm s}$),
the acquisition process moves on to the next bolometer.  This
corresponds to a cycle of 36~periods and results in 2.5~measurements/sec:
\begin{description}
\item[-] The stellar sensor data: \\ 
%%8 ADC values are read at a time and
%%2 bits of each value are encoded into the star sensor word.  
%%The 12 bits of the converter are read in 6 words.  Since we have 6
%%groups of 8 ADCs, we need 36 words to transmit the complete star
%%sensor data;
48~ADC values of 12~bits are read at a time, giving 576~bits. They are
sent during 36~digitisations with 16~bit words.
\item[-] The bolometer measurement after demodulation: the sum of the
bolometer values (16~bits) during one half period (max 48 = 6~bits)
gives a 22~bit number. The LSB bit is lost and 36 words of 21~bits,
corresponding to the 36 measuring channels, are successively sent
during one sampling period (half a modulation period);
\item[-] The synchronisation bits: Two synchronization bits allow data
verification and permit the identification of any problems in the
acquisition chain.  The transputer uses these bits to re--synchronise
the data;
\item[-] The status of valves and heaters: One bit per word gives a
sequence of 72~bits in one period. This sequence encodes the status of
all switches (heaters, valves);
\item[-] The gyros, GPS, temperatures, pressures and flow: during one
half period, this byte contains:
\begin{description}
   \item[-] the even bytes contain one GPS character, if any.
   \item[-] bytes 1 and 3: the X gyroscope data
   \item[-] bytes 5 and 7: the Y gyroscope data
   \item[-] bytes 9 and 11: the Z gyroscope data
   \item[-] bytes 13 and 15: the multiplexed data from
              the dilution board
   \item[-] byte 17: address of the multiplexed data.
\end{description}
\end{description}
The multiplexed data switch between 48 different measurements.

%I.5.2 - Data formatting, compression and telemetry
\subsubsection{Data formatting and compression in the transputer}

The data are transformed by the transputer into blocks of different
types, each block structure containing some partial information.  A
separate task is the transformation of a block into a smaller one
containing the same information after compression.
Tab.~\ref{tab:data_fmt} describes the different block types and gives
their size, percentage of the block emitted, flow rate in kbit/second
and the information content.  These parameters change during
experimental testing and can be modified for flight; the table shows
values appropriate for the Trapani test flight, where the total flow
rate was 84~kbit/second, sent as a bi--phase signal at 108~kbit/second
via telemetry for real--time control. The same data are stored in the
flash EPROM on-board recorder, where a volume of 1~Gbit can be stored
during 24 hours of acquisition.  For the Kiruna flight, the typical
flow rate will correspond to 2.5 blocks per second (modulation 90~Hz).
The total storage required is calculated for 30~hours of observation.
The total compressed data include blocks 12-14 and blocks 1-2-3-4-5
every 4 seconds (1/10 of the blocks).

%I.5.2 - Data flow between the transputer and the telemetry (link2)
\subsubsection{Data flow between the transputer and the telemetry (link2)}

For ground--based testing and calibration, telemetry is passed through
an optical fiber at a rate of 1~Mbit/s. Thus, it is possible to
transmit all information without compression from the experiment.

\section{Ground--based calibration}

Ground tests were planned before the Trapani test flight to
photometrically and optically calibrate the instrument.
Unfortunately, due to the late integration (May 1999) just before the
test flight (July 1999), only methodology could be validated, without
any complete end--to--end results\footnote{The optical ground--based
tests have been obtained since (July 2000)}.  Nevertheless,
performance was judged to be sufficient for the test flight.  Here we
describe the present status of the ground--based calibration devices.

\subsection{Instrument photometric performance}

\begin{figure*}
% as provided by Jean-Charles Vanel 07/03/2000
\resizebox{\hsize}{!}{\includegraphics[angle=0]{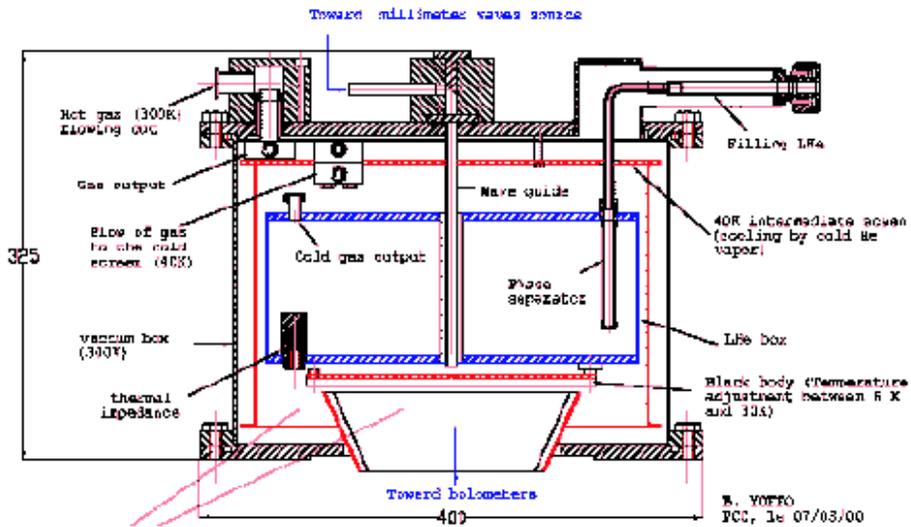}}
\caption{Schematic drawing of the calibration cryostat. This is bolted
to the photometer during the ground--based calibration on top of the
gate valve (Fig.~\ref{fig:cryostat2}. Dimensions are in millimetre.}
\label{fig:calib_cryos}
\end{figure*}

In order to calibrate total transmission, it was decided to use a cold
blackbody (CBB) with a variable temperature in front of the entrance
window of the photometer (with only the thin polypropylene membrane as
a separator). We deduce the total transmission of the photometer by
varying the temperature of the CBB, filling the beams, and measuring
the I-V curves. Here we describe the calibration cryostat that was
designed for this purpose.  A schematic drawing is shown in
Fig.~\ref{fig:calib_cryos}.  The cryostat can be attached even when
the photometer is on the gondola (at an inclination of 28 degrees) to
permit calibration just before flight.  The cryostat is cooled with
liquid Helium only. Helium vapours cool an intermediate screen (IS)
and then go through a heating plate in order to avoid air freezing at
the exit, thereby preventing water from dripping onto the primary
mirror. The CBB plate itself is made of copper covered with a black
material \cite{Lee:2000}.  It is coupled via three copper bands to the liquid
He tank. The blackbody temperature is changed (over a range from 6 and
25 K) via three resistances (with up to 5 Watts of input power).  To
match the various experimental conditions (optical background load due
to various emissivities), the thermal copper links must be
appropriately adapted. Heating the CBB causes more He evaporation,
which then reduces the IS temperature and modifies the optical
background on the 10~K photometer stage.

This calibration cryostat has a built-in light pipe that can feed a
modulated external source through small holes in the CBB, in order to
measure the time constants of the detectors (see
Fig.~\ref{fig:calib_cryos}). This measurement is important for this
kind of scan--modulated CMB experiment, in order to assess the
linearity of the bolometer response as a function of frequency.

The photometer for the Trapani 1999 campaign demonstrated satisfactory
behaviour under the background conditions provided by the calibration
cryostat (at typically 10~K), which were similar to what was expected
for the flight.

No Fourier Transform Spectrometer (FTS) measurements have yet been
performed to check the photometric pixels spectral transmission. For the
interpretation of the test flight data, we have to rely on the
composite transmission as deduced from the single filter and
back--to--back horn transmissions (see Sec.~\ref{ss:FP} and
Fig.~\ref{fig:Arch_band}).

\subsection{Instrument optical calibration}

A pointing table has been built at the ISN laboratory in order to
perform a ground--based determination of the main lobe of the optical
system. The gondola is mounted on this platform. A software has been
developped to pilot the pointing device in order to scan an area
containing a thermal source (placed on a hill at typically 1~km distance).

\begin{figure*}
\resizebox{\hsize}{!}{\includegraphics[angle=270]{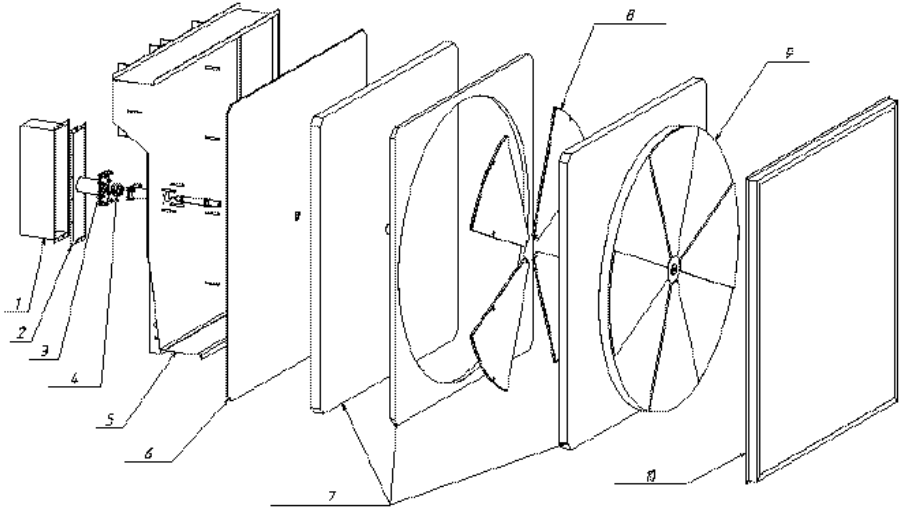}}
\caption{The modulated source in exploded view. 1:~cover, 2:~gasket,
3:~connections, 4:~engine, 5:~box, 6:~Eccosorb background, 7:~heat
insulators (polystyrene), 8:~heating sectors \& Eccosorb, 9:~chopper
(polystyrene wheel + Eccosorb), 10:~cover plate. A detector placed far
above (in this figure) will observe alternating hot (8) and cold (6)
loads as the chopper (9) is rotated.}
\label{fig:ISN_source}
\end{figure*}

For this purpose an extended modulated source ($\simeq 1 \zu m^2$) has
been constructed of which an exploded scheme is shown on
Fig.~\ref{fig:ISN_source}.

This source has a regulated differential temperature of $\Delta T=
50\zu K$. The modulation is triangular. With a typical rotation period
of $0.5\zu s$, the modulated signal has a frequency of 8~Hz (the
source has 4 quadrants: see Fig.~\ref{fig:ISN_source}), which is
easily detected by simple lockin on the bolometer signal with respect
to the thermal background. The source has been tested at the Plateau
de Bure observatory by measuring the main lobe of the POM2 telescope
coupled to the Diabolo cryostat \cite{Benoit:2000}. Some preliminary
measurements were performed at close positions at Trapani just before
the flight.

The very tight schedule before the Trapani flight did not allow 
optimization of all the elements specially conceived for the ground--based
calibration: pointing table, software, extended source, for the
technical flight. Eventually, the optical calibration system
was successfully tested end--to--end in July 2000 and produced
beam sizes of about 8 arcminutes at all frequency bands. This sytem will be
used for the Kiruna flights.  In flight measurements
of the beam sizes during the Trapani campaign are discussed 
in section~\ref{subsec:jup}.

\section{Trapani Test Flight: Description}

%III / Flight description

%III .1  General timing
\subsection{General timing}

Launch of the balloon was provided by ASI (Italian Space Agency) base
near Trapani (Sicily) in the evening of the 17th July 1999 at the end
of the scheduled launch window, in particular when the meteorological
requirements were met.  A logbook summary is given here with hours
given in universal time (decimal hours starting on the 17th) and local
time (hours and minutes) as shown in parenthesis:
\begin{description}
\item[19.37 (21:22)] : launch
\item[21.92 (23:55)] : PID adjustement of the pivot; the gondola
starts to spin at 3~rpm (13.6 deg/sec for a beam at 41~degree
elevation)
\item[21.97 (23:58)] : opening of the protection valve in front of the
membrane (Fig.~\ref{fig:cryostat2}) and the cold optics (this valve
opens automatically when the pressure outside the cryostat drops below
20~mb)
\item[23.00 (01:00)] : Galaxy detected by all bolometers
\item[23.45 (01:27)] : standard altitude achieved; cryostat
temperature stabilizes at 112~mK: final polarisation adjustment for
the bolometers
\item[28.00 (06:00)] : Sunrise
\item[31.53 (09:32)] : protection valve closes automatically when the
temperature of the 10~K stage reaches 13~K due to Sun heating.
\item[34.00 (12:00)] : $^3$He bottle empty
\item[36.32 (14:22)] : separation of the gondola from the balloon.
\item[37.36 (15:23)] : landing in Spain
\item[40.22 (18:13)] : liquid Helium cryostat empty
\end{description}

%III . 2 Characteristics during the flight
\subsection{Flight Characteristics}

%\begin{figure*}
%\resizebox{\hsize}{!}{\includegraphics[angle=90]{balloon_trajectory.ps}}
%\caption{Balloon path over the Mediterranean Sea from launch to landing}
%\label{fig:ball_traj}
%\end{figure*}
%\begin{figure}
%\resizebox{\hsize}{!}{\includegraphics[angle=90]{balloon_altitude.ps}}
%\caption{Balloon altitude (km) as a function of time (hours)}
%\label{fig:ball_alt}
%\end{figure}

The balloon trajectory was nominal for this trans--Mediterranean
flight. After take-off from Trapani (Sicily) the balloon grazed the
African continent before reaching Spain. Altitude was between 40 and
42~km during the useful flight (UT=24 to 36~hours)
%It is described in Fig.~\ref{fig:ball_traj} \& \ref{fig:ball_alt}.

\begin{figure}
%\resizebox{\hsize}{!}{\includegraphics[angle=90]{gren_fig1.ps}}
\resizebox{\hsize}{!}{\includegraphics[angle=90]{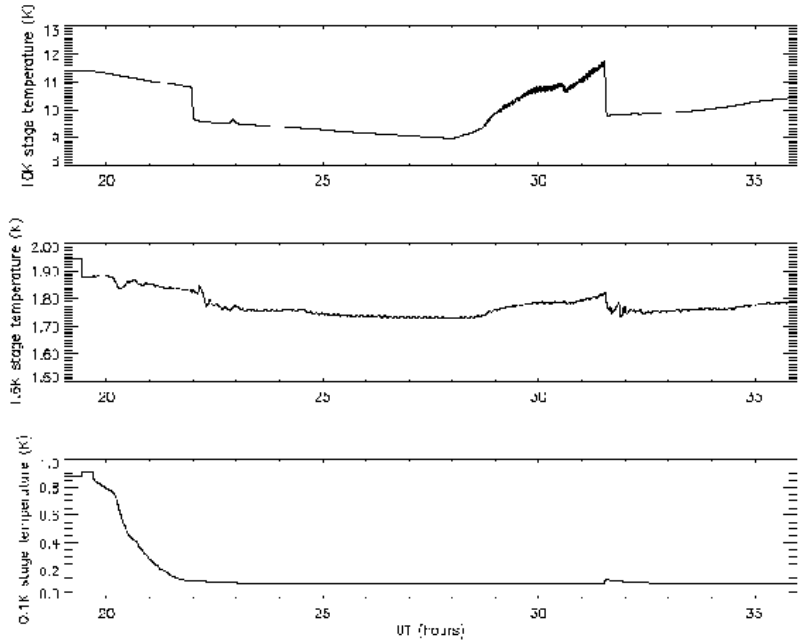}}
% this one is quicker and labelled
\caption{Measured temperature of the different stages as a function of universal time during the
flight. The drop of temperature of the 10~K stage happens when the
valve opens. After sunrise at UT= 28~hr, this stage starts to warm up
until the valve is shut after UT= 31.5~hr.}
\label{fig:temp_sta}
\end{figure}

\begin{figure}
\resizebox{\hsize}{!}{\includegraphics[angle=90]{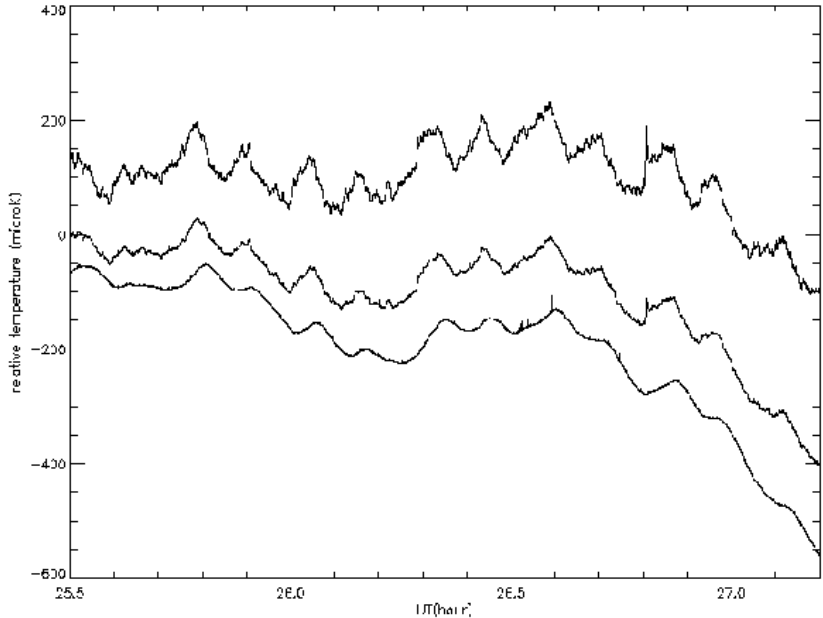}}
\caption{Relative temperature fluctuation of the dilution part of the
cryostat as a fuction of universal time. The top curve is for the
dilution stage, the middle for the intermediate stage and the bottom
curve for the bolometer plate (see Fig.~\ref{fig:th_filtering}).}
\label{fig:temp_100}
\end{figure}

\begin{figure}
\resizebox{\hsize}{!}{\includegraphics[angle=90]{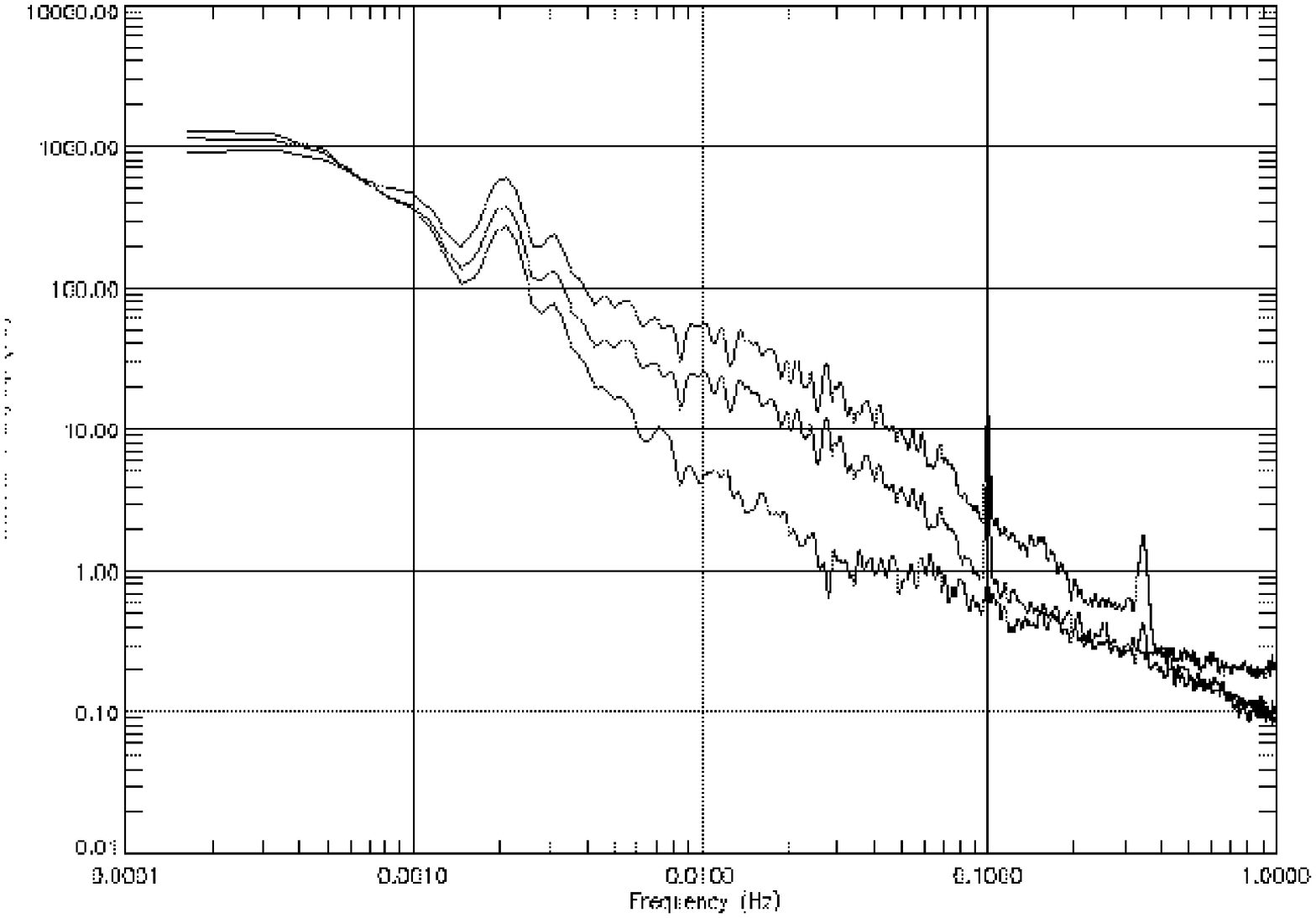}}
\caption{Power spectrum of thermal fluctuations at 100mK. The different
curves represent the fluctuations on the dilution stage, on the
intermediate stage and on the bolometer plate.}
\label{fig:temp_PS}
\end{figure}

Due probably to a small leak in the cryostat, the minimum temperature
reached was only 112~mK and could be obtained only with a higher than
normal flow rate of 24 $\mu$mol/s for $^4$He instead of 16 $\mu$mol/s,
and 8 $\mu$mol/s instead of 4 $\mu$mol/s for $^3$He. The temperatures
of the different stages are represented in Fig.~\ref{fig:temp_sta}:
during the scientific observing time (from 24 to 28~hr.), all
temperatures are relatively stable. The 1.6~K stage fluctuates with an
amplitude of $\sim 10$~mK at frequencies around $2.5\times
10^{-3}\;$Hz; the long--term temperature increase was about 60~mK over
the entire scientific observing time.  The temperature of the 10~K
stage decreased from 10 to 9~K during this period, followed by a
large increase after Sunrise.  The 100~mK stage shows a fluctuation of
1.9~mK (peak to peak).  Temperatures for the different filtering
stages at 100~mK (Fig.~\ref{fig:th_filtering}) are given in
Fig.~\ref{fig:temp_100}, and the fluctuation spectra of these stages
are shown on Fig.~\ref{fig:temp_PS}.  Thermal noise at bolometer level
resides now only at low frequency, while the high frequency white
component is due to thermometer noise.

\section{Trapani Test Flight: in--flight performance}

\subsection{Bolometer signals}

%% Mail Date: Mon, 17 Apr 2000 10:31:55 -0700 (PDT)
%% From: Ravinder Bhatia <rsb@astro.caltech.edu>
%% To: Francois-Xavier DESERT <Francois-Xavier.Desert@obs.ujf-grenoble.fr>
%% I looked at the draft paper and wanted to correct the
%% statement [Section 5.2] that one bolometer died during the flight and that
%% the other was noisy. These statements are incorrect: it was because of
%% Alain's wiring problems that these bolometers failed to deliver good data,
%% not because of the bolometers. If you could correct the paper at your
%% convenience that would be appreciated.
%% Added corrections from Ravinder
%%        Thu, 28 Sep 2000 23:17:51 -0700 (PDT)
%% From:         Ravinder Bhatia <rsb@astro.caltech.edu>

\begin{figure}
\resizebox{\hsize}{!}{\includegraphics[angle=90]{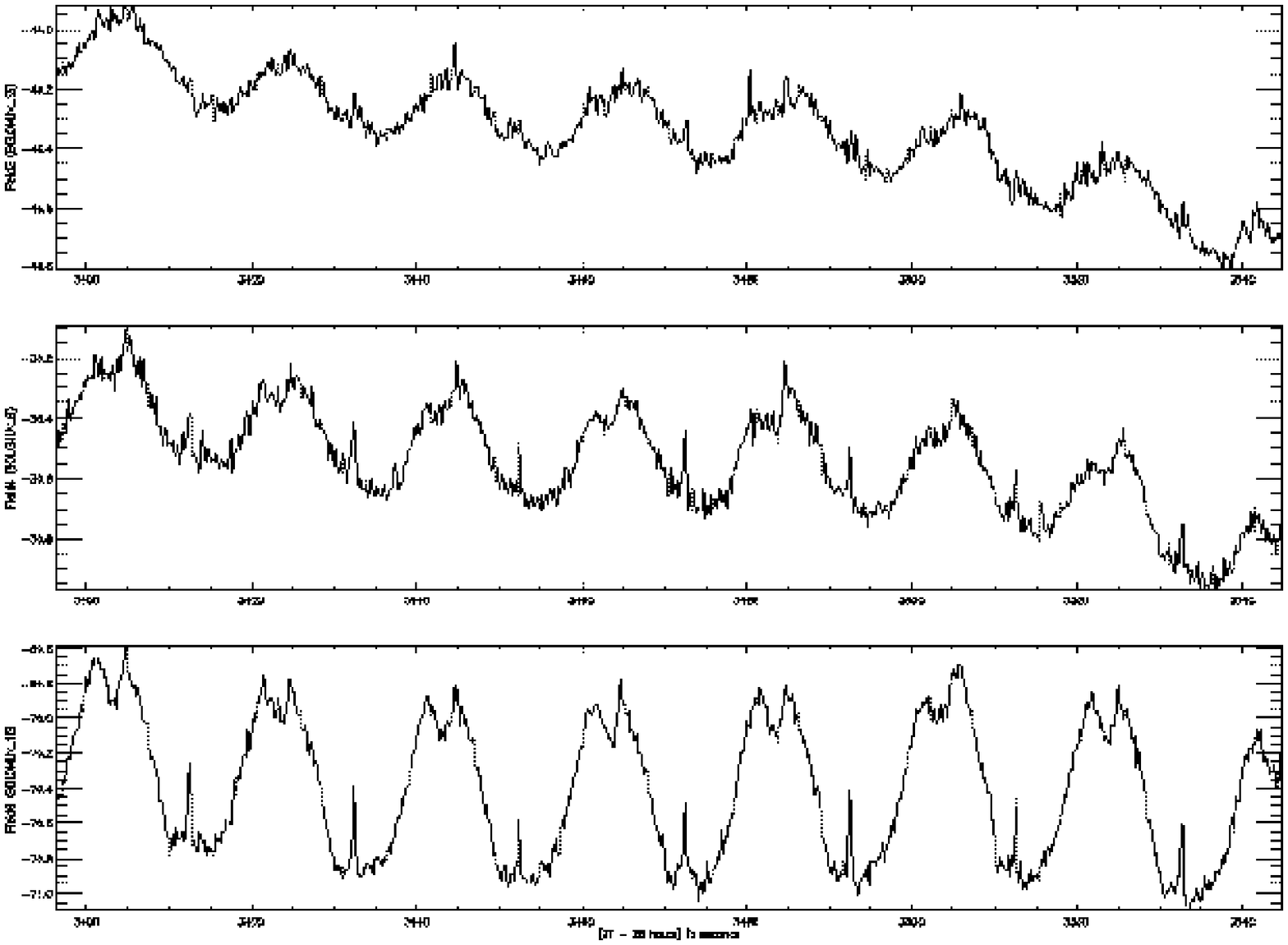}}
\caption{Signal in $\mu$Volts from three bolometers as a function of
time (seconds). From top to bottom, the three channels are shown (143,
217, and 353 GHz) for about 8 revolutions of the gondola. The signal
was deglitched and low-pass filtered ($f_c \simeq 3$~Hz) to show its
broad features.}\label{fig:sig1}
\end{figure}

One of the 143 GHz photometric pixels (B1-2) failed early during the
night and another at 217 GHz (B2-5) was very noisy. Both problems were
due to faulty connections in the electronic system.  The other four
bolometers delivered a continuous and satisfactory signal during the
whole flight. Fig.~\ref{fig:sig1} shows the filtered signal from 3 of
them for few minutes of data. The sharp peaks in the valleys of each
oscillations come from the emission of our galaxy. The CMB dipole is
present at about half the level of the oscillations in the 143~GHz
channels (see subsection~\ref{ss:cmb}).

The main oscillations in the signal in Fig.~\ref{fig:sig1} are due to
a persistent parasitic signal which was present in all bands (with a
high correlation between the bands) during the Trapani flight. It
appears as a strong sine-like oscillation at the spin frequency in the
data and systematically shows two smaller bumps about 70~degrees apart
from the maximum signal. The phase of this signal is not constant
relative to fixed geographical directions or celestial coordinates but
slowly drifts in a non monotonous fashion by about 1.5~turn during the
four hours of night data.  The origin of this signal is not fully
understood.  It is not in phase with variations of the elevation of
the sub-millimeter axis as deduced from preliminary reconstruction
using the stellar sensor information. We therefore can exclude that
this parasitic signal is induced by the swing motion of the
gondola. Intrinsic changes in the atmospheric emission with direction,
such as the presence of clouds with enhanced ozone abundance, can also
be excluded since they would require rather high abundance changes and
a rather unrealistic cloud distribution in order to reproduce the
phase drift observed. A more likely hypothesis is that this signal is
actually due to non-symmetrical emission or reflection of the
atmospheric or earth emission by the balloon, seen through the
far-side lobes of the instrument. This has lead to a new design of the
entrance baffle of the experiment for the forthcoming polar flight
from Kiruna. The charaterisation and removal of this parasitic signal
is one of the main data reduction task ahead.

\subsection{Pointing reconstruction}
\subsubsection{Pointing solution from Galaxy crossings} 

As a first step in the analysis, we obtained an approximate pointing
solution for the experiment using the bright Galactic plane crossing
at each rotation of the gondola. We assumed that there was no
pendulation of the gondola. The IRAS maps were used to compute the
azimuth of the Galactic plane each time it was crossed by the Archeops
353~GHz beam, accounting for the location of the gondola (given by the
GPS).  We could therefore synchronize the azimuth once per rotation
during the flight.  We then interpolated between these Galaxy
crossings in order to obtain an approximate pointing solution. We
estimate the accuracy of this pointing solution to be about 30
arcminutes.

\subsubsection{Pointing reconstruction from the Fast Stellar Sensor}

The final reconstruction of the telescope attitude will be obtained in
three steps: 1)~finding star candidates in the data stream, 2)~associate
these to well-known stars and~3)~relate the solution to the
submillimetre beams. The first step is described in
Sect.~\ref{subs:ss}. The second step has yielded preliminary results
that are used in the following. Final refinements are now being done
and will be reported soon. The third step is described now.

\subsubsection{Focal plane geometry and beams}\label{subsec:jup}
\label{sec:focal_plane_geometry_and_beams}

We made scans of Jupiter when it crossed an elevation of 41 degrees at
%2h48 UT on July 18$^\mathrm{th}$, 1999, when the balloon was
2h48 UT on July 18th, 1999, when the balloon was approximately above
Algiers. Jupiter's azimuth was 110 degrees. Due to our scan strategy,
Jupiter appears as a peak in the bolometer signal at each rotation
when it passes through the focal plane.  Jupiter has an angular
diameter of less than 40~arcseconds and is therefore much smaller than
our beams.  We may therefore consider it as a point source for the
purpose of mapping the beams and reconstructing the focal plane
geometry.

We extract the focal plane geometry by measuring the relative shift in
azimuth and elevation of each bolometer between the observed signal
and the true position of Jupiter on the sky.  The fast star sensor
analysis provided sky coordinates for a point located roughly in the
middle of the focal plane with an accuracy of about 3 arcminutes. The
focal plane geometry is then measured relative to that point for each
bolometer by fitting the location of a 2--dimensional Gaussian to the
image of Jupiter.  In-scan FWHM of 11~arcminutes($\pm 1$) are found
for all bolometers  while cross-scan FWHM of 13~arcminutes are
measured except for bolometer 3-6 (24~arcminutes reflect the
double-peaked structure).
%The positions of the bolometers relative to that
%reference point are given in Table~\ref{tab:foc}.
%\begin{table}
%\caption{Positions of the bolometers relative to the reference point
%(in arcminutes), together with their full--width at half--maximum in
%both scan and cross--scan directions, as determined by a Gaussian fit
%to the beam profiles obtained on Jupiter. Uncertainties are of the
%order of one arcminute. }
%\begin{center}
%\begin{tabular}{|c|cc|cc|}
%\hline
%% Bolometer & $\parallel\mathrm{scan}$ & $\perp\mathrm{scan}$ & FWHM$_{\parallel\mathrm{scan}}$ & FWHM$_{\perp\mathrm{scan}}$\\
%Bolometer & \multicolumn{2}{c}{Beam Location} &
%\multicolumn{2}{c}{Beam Size} \\ \hline & $\parallel\mathrm{scan}$ &
%$\perp\mathrm{scan}$ & FWHM$_{\parallel\mathrm{scan}}$ &
%FWHM$_{\perp\mathrm{scan}}$\\ 
%\hline
%%1-1    &  1.5   & -5.3   & 10.6    & 13.6 \\
%%1-3    &  28.2  & -46.6  & 10.9    & 12.2 \\
%%2-4    & -24.1  &  40.6  & 11.9    & 14.4 \\
%%2-5    &  53.3  & -7.5   & 9.5     & 15.2 \\
%%3-6    & -52.1  & -7.6   & 10.3    & 23.9 \\
%1-1    &  2   &   -5   & 11    & 14 \\
%1-3    &  28  &  -47   & 11    & 12 \\
%2-4    & -24  &   41   & 12    & 14 \\
%2-5    &  53  &   -8   & 10     & 15 \\
%3-6    & -52  &   -8   & 10    & 24 \\
%\hline
%\end{tabular}
%\end{center}
%\label{tab:foc}
%\end{table}

\begin{figure}
\resizebox{\hsize}{!}{\includegraphics[angle=0]{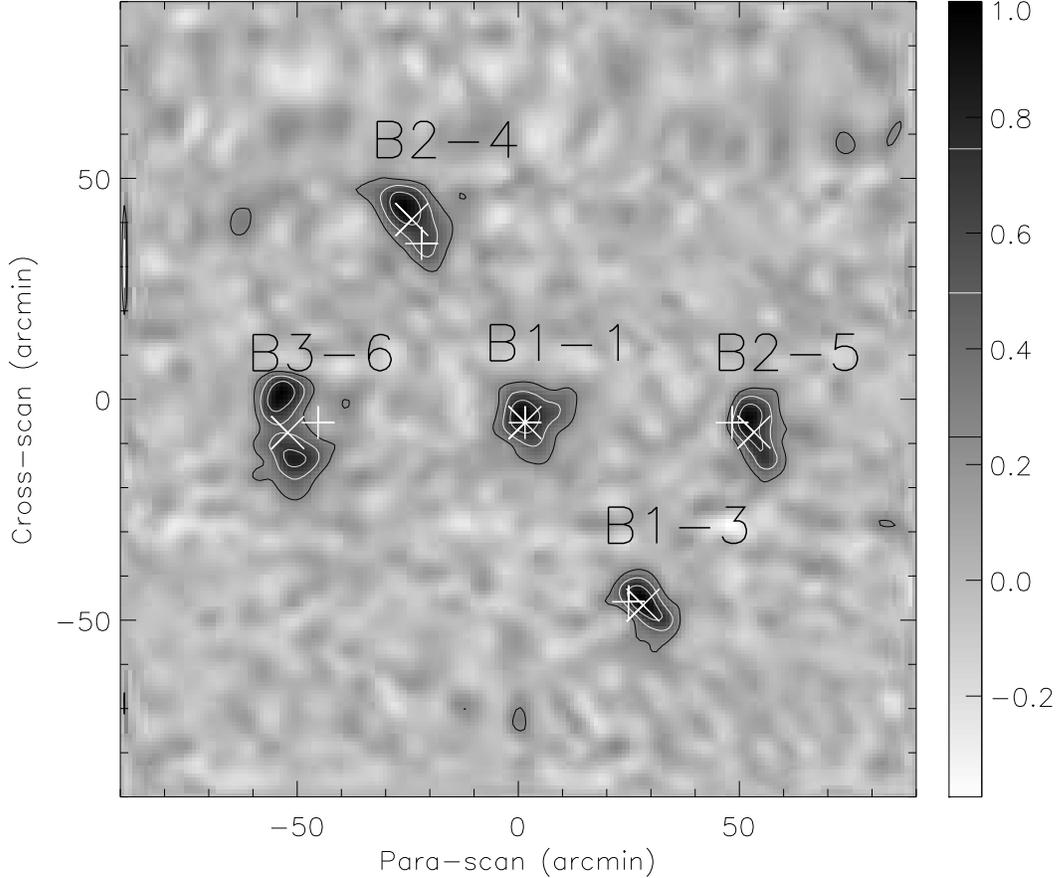}}
\caption{The focal plane configuration during the Archeops flight
in Trapani and the
measured beam shapes. The white pluses show the expected location 
of the photometers relative to the center one. The white crosses are 
the centers of a Gaussian fit to the measured beams (preliminary results).
Photometer B1-2 did not work during the flight and is therefore not
shown in the figure.
}
\label{fig:figfocal}
\end{figure}

The beam maps were obtained by projecting the data from each bolometer
using a preliminary attitude reconstruction from the fast stellar
sensor around the location of Jupiter.  Fig.~\ref{fig:figfocal} shows
the measured beams for our bolometers and their relative positions in
the focal plane. We emphasize that our scan strategy provides only a
few passes across Jupiter for every photometric pixel, which typically
give less than 10 detector samples within a two dimensional full-width
at half-maximum.  Therefore, the beam maps are noisy and are sample
limited, and the reconstructed sizes and shapes should be considered
with caution. The double peaked shape of bolometer 3--6 seems,
however, to be real as it is also observed in the Saturn scans.  A
misalignment in the optical components of the telescope or focal plane
optics could produce the double peaked shape, as well as asymmetry in
the beams.

%\begin{figure}
%\resizebox{\hsize}{!}{\includegraphics[angle=0]{focalplane_nb.epsi}}
%\caption{Measured beam shapes for our 5 bolometers represented at
%their relative position within the focal plane (preliminary
%results). The white pluses show the theoretical pre-flight
%configuration. The white crosses are obtained from the Gaussian fit
%centers.}
%\label{fig:figfocal}
%\end{figure}

\section{In--flight photometric calibration}
We devised 3 schemes to independently calibrate the photometry of the
experiment. The final CMB calibration comes from the CMB dipole
because it has the same spectrum as the CMB anisotropies. Other
methods allow to check the consistency of the calibration accross the
$l$ spectrum and allow the calibration of ``non--CMB'' channels.

\subsection{Photometric calibration on Jupiter}
As Jupiter is much smaller than our beam it can be used as a point
source for calibration. We assume a temperature of
$T^p_{\mathrm{RJ}}=170 \mathrm{K_{RJ}}$ \cite{Goldin:1996}.  The
calibration coefficient is obtained through:

\begin{equation}
\alpha=\frac{T^p_{\mathrm{RJ}}\times\Omega_p}{I_p}
\qquad \left[  \mathrm{mK_{RJ}}.{\mu V}^{-1}\right],
\end{equation}

where $I_p$ is the total measured integrated flux of Jupiter in $\mu
\mathrm{V} .\mathrm{arcmin}^2$ and $\Omega_p$ is the known solid angle
(in $\mathrm{arcmin}^2$) subtended by the planet. The results obtained
with Jupiter for the calibration are shown in Tab.~\ref{tab:cal}.

\subsection{Photometric calibration on the Galaxy}

\begin{figure}
\resizebox{\hsize}{!}{\includegraphics[angle=0]{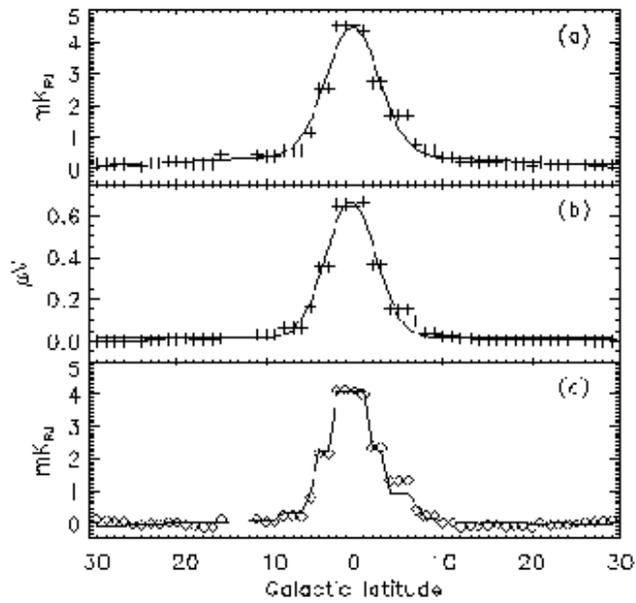}}
\caption{Exemple of cross-calibration between Archeops and FIRAS data
at 353~GHz (at Galactic longitude l=33$^o$).  (a):~FIRAS Galactic
latitude profile (crosses) and its Gaussian fit (continuous line)
(b):~ARCHEOPS Galactic latitude profile (crosses) and its Gaussian fit
(continuous line) (c):~FIRAS Galactic latitude profile (diamonds)
together with the Archeops one (continuous line) multiplied by the
calibration factor of 6.4 mK$_{RJ}/\zu \mu V$.  }
\label{fig:galaxy}
\end{figure}

In--flight calibration of the three Archeops bands was done on the
Galaxy using FIRAS data.  We first constructed maps at 143, 217 and
353~GHz by fitting FIRAS spectra with Planck curves modified by
$\nu^{\alpha}$ emissivity laws.  Archeops Galactic maps where then
convolved with the FIRAS point spread function and color corrected
using the FIRAS--fitted parameters. For selected longitudes with good
signal--to--noise ratios, we computed the Archeops and FIRAS Galactic
profiles in latitude ($-30^o<b<30^o$).  These profiles were fitted by
Gaussian curves, and calibration coefficients obtained by comparing
the heights of these two curves.

At each selected longitude, the shapes of the two profiles show good
agreement (see an example in Fig.~\ref{fig:galaxy}).  There is a small
dispersion in calibration coefficients from point to point.  Derived
values are in Tab.~\ref{tab:cal}.

\begin{table}
\caption{Preliminary calibration values in mK$_{RJ}$/$\mu$V
(error bars will be refined but are below 2~mK$_{RJ}$/$\mu$V)}
\begin{flushleft}
\begin{tabular}{|l|l|l|l|l|}
\hline
bolometer       &  1-1  & 1-3 & 2-4 & 3-6 \\
frequency (GHz) &  143  & 143 & 217 & 353 \\
\hline
Galaxy          &  12.2 & 11.1 & 6.3 & 6.4 \\
Jupiter         &  15.3 & 15.3 & 8.7 & 8.1 \\
Dipole          &  12.2 & 14.7 & 6.7 & --  \\
\hline
\end{tabular}
\end{flushleft}
\label{tab:cal}
\end{table}

\subsection{Photometric calibration on the CMB Dipole}\label{ss:cmb}

Using the 850 micron band (353~GHz) as a template for the parasitic
signal, and assuming that the lower frequency channel contains only a
linear combination of the parasitic and dipole signal (as known from
COBE data) one obtains the dipole calibration shown in
Tab.~\ref{tab:cal}. We note that the dipole calibration accuracy is
well below what can be expected ultimately, probably because it is
still hampered by unidentified signal near the spin frequency in the
data, and is by no means final.

\subsection{Noise}
% Using 10nV/sqrt(Hz) (a bit optimistic: 13 is better)
The final sensitivity of the bolometers during the test flight can be
estimated as (for the white noise at high frequency)
% print, 10.*[13, 6.7, 6.4]
130, 67, 64~$\mu$K$_{RJ}$Hz$^{-1/2}$ equivalent to
% print, 10.*[13, 6.7, 6.4]*[1.72, 2.87, 11.5]
220, 190, 740~$\mu$K$_{CMB}$Hz$^{-1/2}$ at 143, 217, and 353~GHz (2.1,
1.4, and 0.85~mm) respectively. With 3 bolometers on the CMB sensitive
frequencies one can expect an instrument CMB sensitivity of
% print, 1./sqrt( 2./(220.^2)+ 1./(190.)^2)
120~$\mu$K$_{CMB}$Hz$^{-1/2}$ during the 4 hours in dark night. These
sensitivities are very similar to the ones obtained in other
balloon-borne experiments like Boomerang \cite{Bernardis:2000} and
Maxima \cite{Hanany:2000}. The gain brought by the 100~mK stage is
mainly to allow fast bolometers (at most few milliseconds of time
constant). Understanding the full power spectrum of the bolometer
noise is one of the main tasks of the data reduction.

\subsection{Sky coverage}

\begin{figure*}
\resizebox{\hsize}{!}{\includegraphics[angle=90]{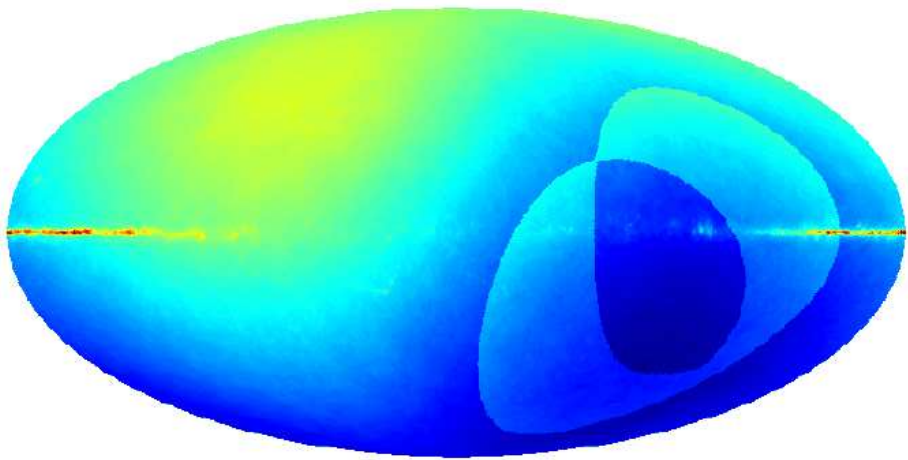}}
\caption{All--sky Mollweide projection of the Galaxy and CMB dipole
emission expected at 143~GHz, centered on the galactic anticenter. The
fraction of the sky covered with the Trapani Flight during the 4 hours
of dark night is shown as a brighter zone superimposed on the map. The
galaxy map is taken from IRAS all-sky survey at 100~\micron
\cite{Schlegel:1998} and extrapolated with a dust spectrum of 17~K and
a $n=2$ emissivity law.}
\label{fig:trap_cov}
\end{figure*}

During the 4 hours of dark night for the Trapani flight (valid for CMB
analysis), the observed sky is as shown in Fig.~\ref{fig:trap_cov}. It
covers about 19\% of the total sky.

\section{Conclusion}

\begin{figure*}
\resizebox{\hsize}{!}{\includegraphics[angle=90]{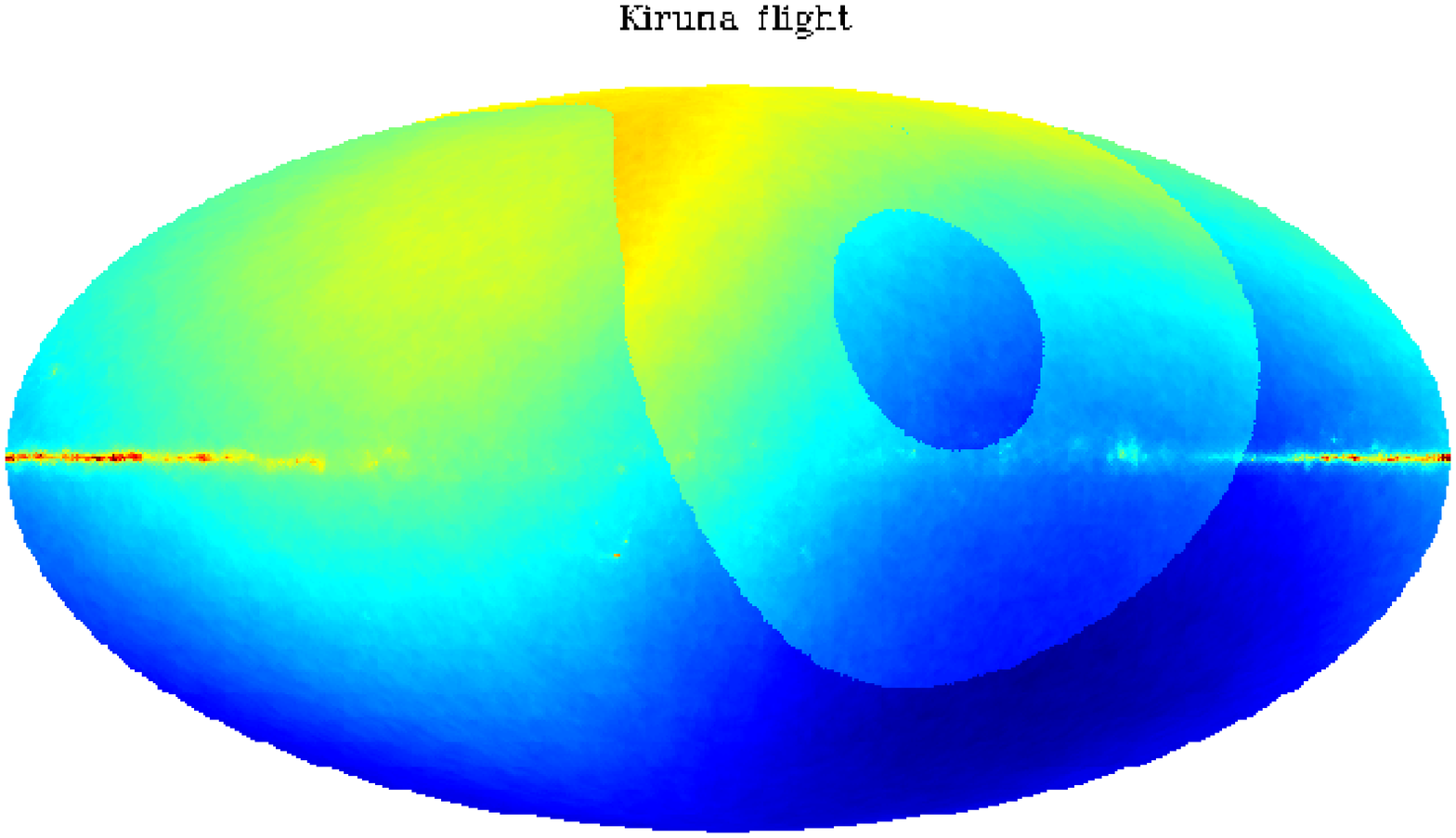}}
\caption{Same simulated all--sky Mollweide projection as previous
figure, of the Galaxy and CMB dipole emission expected at 143~GHz,
centered on the galactic anticenter. The expected fraction (about
28\%) of the sky covered with the next Kiruna flight during the 24
hours is shown as a brighter zone superimposed on the map.}
\label{fig:kir_cov}
\end{figure*}

We have shown the main technical description of the Archeops balloon
experiment. Then we present the status and preliminary results from
the test flight in Trapani July 1999. It is the first time a dilution
fridge has been embarked on a balloon flight. This is an important
result because the dilution cryostat is very similar to the one in
Planck HFI. It is also the first time that 3 independent calibration
methods are used (planet, dipole and Galaxy). They give reasonably
consistent results (Tab.~\ref{tab:cal}). Full data reduction of the
Trapani flight is in progress and should be reported soon. Kiruna
flight \footnote{A successful flight from Kiruna has just happened on
the 29th, January 2001 and should be reported on soon, see
{\tt http://journal.archeops.org/KirunaS1/KirunaS1.html}} is being
prepared using feedback from Trapani flight results. The expected sky
coverage of 28\% is obtained for a nominal 24~hr flight and is shown
in Fig~\ref{fig:kir_cov}.

%________________________________________________________________
%\begin{acknowledgements}
\ack We thank ASI (Italian Space Agency) for its continuous support during
the test flight from Trapani Base. We thank Programme National de
Cosmologie, Centre National d'Etudes Spatiales (CNES) and
participating laboratories for their support during the instrument
design and construction, and DEMIRM--Observatoire de Paris for their
early support.  The experiment benefitted much from the early efforts
by the late Richard Gispert to help us building a robust data
reduction pipeline. We thank K. Gorski for providing us the Healpix
software package.

%\end{acknowledgements}

\end{document}